%% file: paper.tex
\pdfoutput=1

\documentclass[10pt, journal, comsoc]{IEEEtran}
\usepackage[T1]{fontenc}
\usepackage{textcomp}
\usepackage{booktabs}
\usepackage[pdftex]{graphicx}
\usepackage{url}
\usepackage{acronym} \input{acronyms}
\usepackage{tablefootnote}
\usepackage{xcolor}
\usepackage{color, colortbl}
\definecolor{Gray}{gray}{0.9}
\usepackage{multirow,array}
\usepackage{makecell}
\usepackage{tabularx}
\usepackage{amsmath, amssymb}
\usepackage[ruled,linesnumbered,vlined]{algorithm2e}
\SetAlFnt{\small}
\SetAlCapFnt{\small}
\SetAlCapNameFnt{\small}
\usepackage{bbding}

\usepackage{subcaption}

\usepackage{listings}

\definecolor{delim}{RGB}{20,105,176}
\definecolor{numb}{RGB}{106, 109, 32}
\definecolor{string}{rgb}{0.64,0.08,0.08}

\lstdefinelanguage{json}{
    escapechar=\%,
    numbers=left,
    numberstyle=\small,
    frame=single,
    frameround={t}{t}{t}{t},
    captionpos=b,
    rulecolor=\color{black},
    showspaces=false,
    showstringspaces=false,
    showtabs=false,
    breaklines=true,
    postbreak=\raisebox{0ex}[0ex][0ex]{\ensuremath{\color{gray}\hookrightarrow\space}},
    breakatwhitespace=true,
    basicstyle=\ttfamily\small,
    upquote=true,
    morestring=[b]",
    stringstyle=\color{string},
    literate=
     *{0}{{{\color{numb}0}}}{1}
      {1}{{{\color{numb}1}}}{1}
      {2}{{{\color{numb}2}}}{1}
      {3}{{{\color{numb}3}}}{1}
      {4}{{{\color{numb}4}}}{1}
      {5}{{{\color{numb}5}}}{1}
      {6}{{{\color{numb}6}}}{1}
      {7}{{{\color{numb}7}}}{1}
      {8}{{{\color{numb}8}}}{1}
      {9}{{{\color{numb}9}}}{1}
      {\{}{{{\color{delim}{\{}}}}{1}
      {\}}{{{\color{delim}{\}}}}}{1}
      {[}{{{\color{delim}{[}}}}{1}
      {]}{{{\color{delim}{]}}}}{1},
}

\usepackage[backref=false,bookmarks=false,nolinks=true]{hyperref}

\usepackage{orcidlink}

\usepackage{caption}

\usepackage{units}

\usepackage{soul}

\definecolor{LightBlue}{HTML}{CFE2F3}
\definecolor{DarkBlue}{HTML}{6FA8DC}

\usepackage{tikz}
\newcommand{\EmptyCircle}{%
  \tikz[baseline=-0.6ex]\draw[line width=0.4pt] (0,0) circle (0.8ex);%
}
\newcommand{\HalfCircle}{%
  \tikz[baseline=-0.6ex]{
    \draw[line width=0.4pt] (0,0) circle (0.8ex);
    \fill (0,0) -- (0:0.8ex) arc (0:180:0.8ex) -- cycle;
  }%
}
\newcommand{\FullCircle}{%
  \tikz[baseline=-0.6ex]\fill (0,0) circle (0.8ex);%
}

\usepackage{eso-pic}
\AddToShipoutPictureBG{%
  \AtPageUpperLeft{%
    \put(0,\LenToUnit{-11pt}){%
      \makebox[\paperwidth][c]{%
        \parbox[t]{0.92\paperwidth}{\centering\sffamily\fontsize{6.5}{8}\selectfont
          This article has been accepted for publication in IEEE Transactions on Network and Service Management. This is the author's version which has not been fully edited and\\
          content may change prior to final publication. Citation information: DOI 10.1109/TNSM.2026.3751565%
        }%
      }%
    }%
  }%
  \AtPageLowerLeft{%
    \put(0,\LenToUnit{20pt}){%
      \makebox[\paperwidth][c]{%
        \sffamily\fontsize{6.5}{8}\selectfont
        This work is licensed under a Creative Commons Attribution 4.0 License. For more information, see https://creativecommons.org/licenses/by/4.0/%
      }%
    }%
  }%
}

\makeatletter
\ifCLASSINFOpdf
\fi
\makeatother

\makeatletter
\renewcommand{\section}{\@startsection{section}{1}{\z@}{1.8ex plus 1.0ex minus 1.0ex}%
{0.1ex plus 0.5ex minus 0ex}{\normalfont\normalsize\centering\scshape}}
\renewcommand{\subsection}{\@startsection{subsection}{2}{\z@}{2.0ex plus 1.0ex minus 1.0ex}%
{0.1ex plus 0.3ex minus 0ex}{\normalfont\normalsize\itshape}}
\renewcommand{\subsubsection}{\@startsection{subsubsection}{3}{\z@}{1.0ex plus 1.0ex minus 0.5ex}%
{0.1ex plus 0.3ex minus 0ex}{\normalfont\normalsize\itshape}}
\renewenvironment{IEEEbiographynophoto}[1]{%
  \if@IEEEbiographyTOCentrynotmade%
    \setcounter{IEEEbiography}{-1}%
    \refstepcounter{IEEEbiography}%
    \addcontentsline{toc}{section}{Biographies}%
    \global\@IEEEbiographyTOCentrynotmadefalse%
  \fi%
  \refstepcounter{IEEEbiography}%
  \addcontentsline{toc}{subsection}{#1}%
  \normalfont\@IEEEcompsoconly{\sffamily}\footnotesize\interlinepenalty500%
  \vskip 0.5\baselineskip plus 0.2fil minus 0.1\baselineskip%
  \parskip=0pt\par%
  \noindent\textbf{#1\ }\@IEEEgobbleleadPARNLSP}{\relax\par\normalfont}
\makeatother

\ifCLASSINFOpdf
\else
\fi
\begin{document}
\bstctlcite{IEEEexample:BSTcontrol}


\title{Towards Energy- and QoS-aware Load Balancing for 5G Advanced: Leveraging O-RAN to Achieve Sustainability and Energy Efficiency}

%
%
%

\author{Gustavo~Z.~Bruno\orcidlink{0000-0002-1424-3404}, 
Gabriel~M.~Almeida\orcidlink{0000-0002-3764-2336}, 
Aloízio~Da~Silva\orcidlink{0000-0002-3922-9916}, 
Luiz~A.~DaSilva\orcidlink{0000-0001-6310-6150},~\textit{Fellow,~IEEE}, 
Joao~F.~Santos\orcidlink{0000-0001-6439-2056},~\textit{Senior Member,~IEEE}, Alexandre~Huff\orcidlink{0000-0003-0371-4837}, 
Kleber~V.~Cardoso\orcidlink{0000-0001-5152-5323}, and
Cristiano~B.~Both\orcidlink{0000-0002-9776-4888}


\IEEEcompsocitemizethanks{
\IEEEcompsocthanksitem IEEE TNSM, Early Access, 2026. DOI: 10.1109/TNSM.2026.3751565.
\IEEEcompsocthanksitem Gustavo Z. Bruno is with INATEL. E-mail: gustavo.zanatta@posdoc.inatel.br.
\IEEEcompsocthanksitem Cristiano B. Both is with UNISINOS. E-mail: cbboth@unisinos.br.
\IEEEcompsocthanksitem Gabriel M. Almeida and Kleber V. Cardoso are with UFG. E-mails: \{gabrielmatheus, kleber\}@inf.ufg.br.
\IEEEcompsocthanksitem Alexandre Huff is with UTFPR. E-mail: alexandrehuff@utfpr.edu.br.
\IEEEcompsocthanksitem Aloízio Da Silva, Luiz A. DaSilva, and Joao F. Santos are with CCI and VT. E-mails: \{aloiziops, ldasilva, joaosantos\}@vt.edu.
}
}
%
%

\markboth{IEEE Transactions on Network and Service Management (TNSM), Early Access, 2026}%
{Bruno \MakeLowercase{\textit{et al.}}: Towards Energy- and QoS-aware Load Balancing for 5G Advanced}
%



\maketitle

\begin{abstract}
\input{Sections/1-Abstract}
\end{abstract}

\acresetall

\begin{IEEEkeywords}
O-RAN, RIC, 5G Advanced, Energy Efficiency, Dynamic Demand, Resource Management
\end{IEEEkeywords}

%
\IEEEpeerreviewmaketitle

\input{Sections/2-Introduction}
\input{Sections/3-Background}
\input{Sections/4-ProblemFormulation}
\input{Sections/5-Architecture}
\input{Sections/6-Prototype}
\input{Sections/7-Evaluation}
\input{Sections/8-RelatedWorks}
\input{Sections/9-Conclusion}
\input{Sections/10-Acknowledgment}



%
\bibliographystyle{IEEEtran}
\bibliography{IEEEabrv,references}

%
\input{Sections/11-Biography}




\end{document}

%% file: acronyms.tex
\acrodef{3GPP}{3rd Generation Partnership Project}
\acrodef{5G}{Fifth Generation}
\acrodef{6G}{Sixth Generation}
\acrodef{A1}{A1 Interface}
\acrodef{AI}{Artificial Intelligence}
\acrodef{ARQ}{Automatic Repeat Request}
\acrodef{BBU}{Baseband Unit}
\acrodef{BH}{Backhaul}
\acrodef{BLER}{Block Error Rate}
\acrodef{BS}{Base Station}
\acrodef{CDF}{Cumulative Distribution Function}
\acrodef{CNN}{Convolutional Neural Networks}
\acrodef{COTS}{Commercial Off-The-Shelf}
\acrodef{CQI}{Channel Quality Index}
\acrodef{CP}{Control Plane}
\acrodef{CPU}{Central Processing Unit}
\acrodef{CU}{Central Units}
\acrodef{CWDM}{Coarse Wavelength Division Multiplexing}
\acrodef{C-RAN}{Cloud Radio Access Network}
\acrodef{CN}{Core Network}
\acrodef{DCDPandUA}{Dynamic CU and DU Placement and User Association}
\acrodef{Dis}{Disaggregated}
\acrodef{Dis-RIC}{Disaggregated near-RT RIC}
\acrodef{DL}{Downlink}
\acrodef{DP}{Data Plane}
\acrodef{DRAandNFP}{Dynamic Resource Allocation and Network Function Placement}
\acrodef{D-RAN}{Distributed Radio Access Network}
\acrodef{DU}{Distributed Unit}
\acrodef{E2AP}{E2 Application Protocol}
\acrodef{E2Sim}{E2 Simulator}
\acrodef{E2SM}{E2 Service Model}
\acrodef{EPRE}{Energy Per Resource Element}
\acrodef{eNB}{Evolved Node B}
\acrodef{EPC}{Evolved Packet Core}
\acrodef{FCAPS}{Fault, Configuration, Accounting, Performance, and Security}
\acrodef{FH}{Fronthaul}
\acrodef{FL}{Federated Learning}
\acrodef{FlexRIC}{Flexible RAN Intelligent Controller}
\acrodef{LOS}{Line-of-Sight}
\acrodef{gNB}{Next Generation Node B}
\acrodef{gNB-CU}{Next-Generation NodeB - Central Unit}
\acrodef{gNB-DU}{Next-Generation NodeB - Distributed Unit}
\acrodef{GoB}{Grid of Beams}
\acrodef{GPP}{General-Purpose Processor}
\acrodef{HARQ}{Hybrid Automatic Repeat Request}
\acrodef{IoT}{Internet of Things}
\acrodef{IIoT}{Industrial Internet of Things}
\acrodef{IL}{Internal layer}
\acrodef{INFOCOM}{IEEE International Conference on Computer Communications}
\acrodef{IOEE}{Indirect Optimization of Energy Efficiency}
\acrodef{JISA}{Journal of Internet Services and Applications}
\acrodef{JSON}{JavaScript Object Notation}
\acrodef{JSAC}{IEEE Journal on Selected Areas in Communications}
\acrodef{K8s}{Kubernetes}
\acrodef{KPI}{Key Performance Indicator}
\acrodef{LTE}{Long-Term Evolution}
\acrodef{MAC}{Media Access Control}
\acrodef{MC}{Monolithic Co-located}
\acrodef{MC-RIC}{Monolithic Co-located near-RT RIC}
\acrodef{MD}{Monolithic Distributed}
\acrodef{MD-RIC}{Monolithic Distributed near-RT RIC}
\acrodef{MEC}{Multi-access Edge Computing}
\acrodef{MH}{Midhaul}
\acrodef{MIMO}{Multiple-Input, Multiple-Output}
\acrodef{ML}{Machine Learning}
\acrodef{MLB}{Mobility Load Balancing}
\acrodef{mMIMO}{Massive MIMO}
\acrodef{MRO}{Mobility Robustness Optimization}
\acrodef{Near-RT}{Near Real-Time}
\acrodef{Near-RT RIC}{Near Real-Time RIC}
\acrodef{NFV}{Network Function Virtualization}
\acrodef{NFVI}{Network Function Virtualization Infrastructure}
\acrodef{NG-Core}{Next Generation Core}
\acrodef{NG-RAN}{Next-Generation Radio Access Network}
\acrodef{ng-eNB}{Next-Generation - Evolved NodeB}
\acrodef{NIC}{Network Interface Card}
\acrodef{NIB}{Network Information Base}
\acrodef{Non-RT}{Non Real-Time}
\acrodef{Non-RT RIC}{Non Real-Time RIC}
\acrodef{NSI}{Network Slice Instance}
\acrodef{NSSI}{Network Slice Subnet Instance}
\acrodef{O-CU}{Open Central Unit}
\acrodef{O-DU}{Open Distributed Unit}
\acrodef{O-Cloud}{Open Cloud}
\acrodef{OAMF}{\ac{O-RAN} Adaptive Management Framework}
\acrodef{OPlaceRAN}{Orchestrator Placement for Radio Access Network}
\acrodef{O-RAN}{Open Radio Access Network}
\acrodef{OSC}{\ac{O-RAN} Software Community}
\acrodef{O-RU}{Open Radio Unit}
\acrodef{P2P}{Point-To-Point}
\acrodef{PDCP}{Packet Data Control Protocol}
\acrodef{PHY}{Physical Layer}
\acrodef{PON}{Passive Optical Network}
\acrodef{QAM}{Quadrature Amplitude Modulation}
\acrodef{QoS}{Quality of Service}
\acrodef{QoE}{Quality of Experience}
\acrodef{RAM}{Random Access Memory}
\acrodef{RAN}{Radio Access Network}
\acrodef{RF}{Radio Frequency}
\acrodef{RFSS}{RF Stadium Simulator}
\acrodef{RIC}{RAN Intelligent Controller}
\acrodef{RIC-O}{RIC Orchestrator}
\acrodef{RLC}{Radio Link Control}
\acrodef{RMR}{RIC Message Router}
\acrodef{RRC}{Radio Resource Control}
\acrodef{RRH}{Remote Radio Head}
\acrodef{RRM}{Radio Resource Management}
\acrodef{RSRP}{Reference Signal Received Power}
\acrodef{RSRQ}{Reference Signal Received Quality}
\acrodef{RSSI}{Received Signal Strength Indicator}
\acrodef{RU}{Radio Units}
\acrodef{SCA}{Successive Convex Approximation}
\acrodef{SCTP}{Stream Control Transmission Protocol}
\acrodef{SDL}{Shared Data Layer}
\acrodef{SDN}{Software-Defined Networking}
\acrodef{SDR}{Software Defined Radio}
\acrodef{SDAP}{Service Data Adaptation Protocol}
\acrodef{SD-RAN}{Software-Defined Radio Access Network}
\acrodef{SBRC}{Simpósio Brasileiro de Redes de Computadores e Sistemas Distribuídos}
\acrodef{SBRT}{Simpósio Brasileiro de Telecomunicações}
\acrodef{SINR}{Signal-to-Interference-plus-Noise Ratio}
\acrodef{SLA}{Service Level Agreement}
\acrodef{SM}{Service Model}
\acrodef{SMO}{Service Management and Orchestration}
\acrodef{SNR}{Signal-to-Noise Ratio}
\acrodef{srsRAN}{Software Radio Systems RAN}
\acrodef{STSL}{Shared Time-Series Layer}
\acrodef{TDR}{Time to Disaster Recovery}
\acrodef{TDD}{Time Division Duplex}
\acrodef{TNSM}{IEEE Transactions on Network and Service Management}
\acrodef{UAV}{Unmanned Aerial Vehicle}
\acrodef{UE}{User Equipment}
\acrodef{UL}{Uplink}
\acrodef{UP}{User Plane}
\acrodef{USRP}{Universal Software Radio Peripheral}
\acrodef{VES}{\ac{VNF} Event Streaming}
\acrodef{VESPA}{\ac{VES} Prometheus Adapter}
\acrodef{vCPU}{Virtual Central Processing Unit}
\acrodef{vCU}{Virtualized Central Unit}
\acrodef{vDU}{Virtualized Distributed Unit}
\acrodef{vNG-RAN}{virtualized Next Generation Radio Access Network}
\acrodef{vRAN}{Virtualized Radio Access Network}
\acrodef{vRU}{Virtualized Radio Unit}
\acrodef{VM}{Virtual Machine}
\acrodef{VNF}{Virtual Network Function}
\acrodef{Y1}{Y1 Interface}
\acrodef{O1}{O1 Interface}
\acrodef{O2}{O2 Interface}
\acrodef{R1}{R1 Interface}
\acrodef{O-RAN SMO}{\ac{O-RAN} Service Management and Orchestration}
\acrodef{O-FH}{Open Fronthaul}
\acrodef{MILP}{Mixed Integer Linear Programming}
\acrodef{REST}{Representational State Transfer}
\acrodef{API}{Application Programming Interface}
\acrodef{URLLC}{Ultra-Reliable Low-Latency Communication}
\acrodef{E2}{E2 Interface}
\acrodef{E2SM-KPM}{E2 Service Model for Key Performance Measurement}
\acrodef{E2SM-RC}{E2 Service Model for Radio Control}
\acrodef{EQLite}{Energy- and QoS-aware Load-balancing ITErative heuristic}
\acrodef{MCC}{Mobile Country Code}
\acrodef{MNC}{Mobile Network Code}
\acrodef{PCI}{Physical Cell Identifier}
\acrodef{IMSI}{International Mobile Subscriber Identity}
\acrodef{EIRP}{Effective Isotropic Radiated Power}
\acrodef{UHD}{Ultra High Definition}

%% file: Sections/1-Abstract.tex
The increasing energy consumption of next-generation mobile networks necessitates the adoption of autonomous and energy-aware management strategies.
This article proposes a novel adaptive solution leveraging the Open Radio Access Network (O-RAN) architecture to optimize energy efficiency while maintaining Quality of Service (QoS).
The proposed approach introduces a hierarchical O-RAN-aligned control framework in which a Non-Real-Time RAN Intelligent Controller (Non-RT RIC) periodically computes energy-aware policies from long-term historical data, while a Near-Real-Time RIC enforces those policies through per-User Equipment (UE) handover actions on a sub-second timescale.
We formulate a joint energy- and QoS-aware load balancing problem as a Mixed-Integer Non-Linear Programming (MINLP) model.
This model optimizes UE association across E2 Nodes to minimize transmission power and autonomously deactivate underutilized cells, while satisfying strict QoS constraints.
To validate the proposed solution, we deploy it in an experimental environment that simulates massive sports events.
Experimental results demonstrate 72\% energy savings over a 24-hour trace and sub-second handover delays (0.123--0.204\,s/UE), confirming the solution's feasibility for autonomous and sustainable 5G Advanced networks.
This work provides guidelines for deploying energy-efficient strategies in O-RAN environments and underscores the potential of adaptive solutions for sustainable mobile communications.

%% file: Sections/2-Introduction.tex
\section{Introduction} \label{sec:introduction}

While cellular generations improve data rates through advanced signal processing and denser deployments, they significantly increase infrastructure power consumption~\cite{Kumar2023}.
Mitigating this environmental impact is a key priority for 5G Advanced sustainability initiatives~\cite{Gonçalves2020, Lopez-Perez2022}.
However, implementing energy-saving strategies in traditional, proprietary network architectures remains challenging due to a lack of flexibility and open interfaces, creating barriers to practical deployment and validation.

Among future network initiatives, the O-RAN ALLIANCE~\cite{O-RAN-Alliance} envisions extending \ac{5G} by embedding programmability and intelligence in the \ac{RAN}~\cite{Almeida2024}. It enables mobile network operators to customize their networks with data-driven control loops that target diverse optimization objectives, from enhancing data rates to reducing power consumption. In addition to providing architectural components and interface definitions, the \ac{OSC} offers reference implementations of \ac{O-RAN} entities that adhere to the \ac{O-RAN} specifications, serving as a prototyping baseline. Moreover, \ac{O-RAN} standardizes the deployment of custom data-driven control loops through xApps and rApps (detailed later in Section~\ref{sec:O-RAN-architecture}). Therefore, \ac{O-RAN} stands out as a pivotal platform for the development, validation, and dissemination of practical energy saving solutions.

The O-RAN ALLIANCE recognizes energy efficiency as a key area of focus in its specifications, which has inspired a range of studies and use cases~\cite{O-RAN.WG1.NESUC}. Although several works~\cite{Hoffmann2024, Salvat2023, Hammami2022, Wang2022, Abedin2022} have explored energy saving strategies leveraging the disaggregated network architecture inherent in \ac{O-RAN}, much of the current literature remains predominantly focused on analytical models and simulation-based evaluations. More recent studies (e.g.,~\cite{Abubakar2023, Rimedo2023}) have begun to address practical aspects and experimental validations, while forward-looking initiatives aimed at aligning \ac{O-RAN} with sustainable 5G Advanced networks have been detailed by the O-RAN ALLIANCE~\cite{O-RAN-Alliance-6G-2023}. Building on these advances, our approach distributes tasks related to monitoring, control actions, and \ac{UE} handover decisions across different \ac{O-RAN} entities. This practical focus reinforces \ac{O-RAN}'s potential as a robust platform for deploying energy-saving solutions in 5G Advanced networks.

This article introduces the first approach to energy-efficient communication towards 5G Advanced, leveraging a practical \ac{O-RAN} environment. First, we formulate an energy- and QoS-aware load-balancing problem that dynamically reallocates \acp{UE} among neighboring cells, ensuring \ac{QoS} and minimizing O-RU transmit power. We also consider powering off certain E2 Nodes to further conserve energy. The objective is to maximize network efficiency by balancing mobility load and energy savings while managing the resulting trade-off with user performance. Moreover, we implement a data-driven control loop in a practical \ac{O-RAN} environment, deploying the solutions from our optimization model to perform handovers, reduce the number of active E2 Nodes, and lower energy consumption while maintaining connectivity and balancing \ac{QoS} trade-offs. The main contributions of this article are as follows:

\begin{itemize}
    \item \textbf{Joint optimization for energy-aware reassociation:} We formulate a joint optimization problem for energy-efficient reassociation of \acp{UE} across neighboring \acp{O-RU} with explicit \ac{QoS} and mobility load-balancing constraints. To efficiently solve this problem in practical, large-scale scenarios, we introduce EQLite, a scalable heuristic algorithm. The model captures both \ac{O-RU} on/off decisions and transmit-power adaptation to minimize active infrastructure and radio power while preserving \ac{QoS}. This unified treatment of energy saving and mobility management is detailed in Section~\ref{sec:model}.
    \item \textbf{\ac{O-RAN}-compliant closed-loop design:} We propose a hierarchical two-tier control architecture aligned with \ac{O-RAN}. The Energy Savings rApp, hosted in the \ac{Non-RT RIC}, periodically computes energy-aware policies from long-term historical data on a timescale of minutes, consistent with the \ac{Non-RT RIC}'s $>$1\,s control-loop specification. A Handover xApp in the \ac{Near-RT RIC} enforces these policies by issuing per-\ac{UE} handover commands on a millisecond-to-second timescale via the \ac{E2} interface. The rApp computes policies from the optimization outputs and delivers them over A1, while the xApps enforce decisions on E2 and interact with the RAN in near real-time to execute energy-aware handovers. The architecture and interfaces are described in Section~\ref{sec:architectural-components}.
    \item \textbf{Experimental validation on a realistic \ac{O-RAN} testbed:} We prototype the full loop on an \ac{OSC}-based \ac{O-RAN} testbed with standard interfaces and trace-driven traffic patterns, using an emulated \ac{RAN} (\ac{RFSS}/\ac{E2Sim}) that exercises the \ac{E2AP}/\ac{E2SM-KPM}/\ac{E2SM-RC} signaling plane over production \ac{O-RAN} interfaces, including a stadium scenario with up to 8{,}192 \acp{UE} and 56~\acp{O-RU}. Over a 24-hour trace, the adaptive policy reduces site energy from $\approx\!65.4$\,kWh (all-on baseline) to $\approx\!18.3$\,kWh, yielding $72.08\%$ energy savings, while managing \ac{QoS} trade-offs as detailed in the evaluation. Off-peak, the controller maintains $3$--$6$ active O-RUs versus $\sim\!35$--$37$ at peak load (Fig.~\ref{fig:eirp-vs-ue}); fine-grained steps achieve instantaneous savings of $\sim\!24\%$ at nearly constant load (Fig.~\ref{fig:eirp-vs-ue-zoom}). Handover latency remains below $1$\,s per \ac{UE} across 256--8{,}192 \acp{UE} (e.g., $0.123$--$0.204$\,s/\ac{UE}). Results are presented in Section~\ref{sec:energy-results}.
    \item \textbf{System extensions enabling energy-aware operation:} We extend key \ac{O-RAN} components, such as the \ac{VES} Collector, \ac{VESPA} Manager, A1 Mediator, and E2Sim. We developed \iac{RFSS}, a Monitoring xApp, and a Handover xApp to support energy savings. Our contributions include end-to-end event stream integration (validated with the full \ac{VES}\slash{}\ac{VESPA} pipeline up to 1{,}024~\acp{UE}; a simplified Prometheus path is used at larger scales), policy instance management and tracking, persistent state in \ac{K8s}, realistic \ac{UE} behavior and handover simulation, and fine-grained monitoring of computing resource utilization. Implementation details appear in Section~\ref{sec:prototype}. To ensure replicability, the source code is available at \url{https://github.com/zanattabruno/Energy-Saver-Tests}.
\end{itemize}

The remainder of this article is organized as follows. Section~\ref{sec:O-RAN-architecture} provides background on \ac{O-RAN} and its role in adaptive network management for 5G Advanced. Section~\ref{sec:model} formulates the optimization problem for energy-efficient \ac{UE} reassociation and power-saving strategies. Section~\ref{sec:architectural-components} details the architecture of our proposed \ac{O-RAN}-based approach, with emphasis on xApps and rApps. Section~\ref{sec:prototype} describes the practical implementation and modifications made to the \ac{O-RAN} platform to enable our energy saving solution. Section~\ref{sec:energy-results} presents experimental results from our \ac{O-RAN} environment, and Section~\ref{sec:related-works} reviews related work, highlighting gaps in existing energy-efficient solutions for 5G Advanced networks. Finally, Section~\ref{sec:conclusion} concludes the article and outlines directions for future research.

%% file: Sections/3-Background.tex
\section{Technical Background in O-RAN} \label{sec:O-RAN-architecture}

\begin{figure}[t]
\centering
\includegraphics[width=0.75\linewidth]{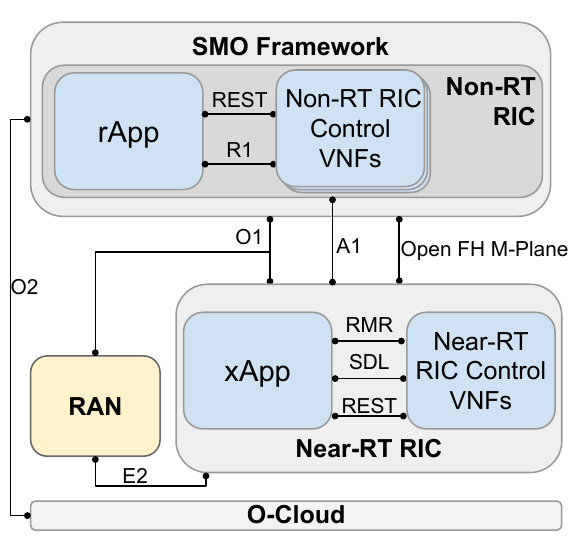}
\caption{Key O-RAN interfaces: SMO, Non-RT RIC, Near-RT RIC, and RAN.}
\label{fig:ric_architecture}
\end{figure}

The \ac{O-RAN} architecture, developed by the O-RAN ALLIANCE, introduces programmability and open interfaces to cellular networks.
As shown in Fig.~\ref{fig:ric_architecture}, it consists of the \ac{SMO} framework, the \ac{Non-RT RIC} (hosting rApps for long-term optimization), and the \ac{Near-RT RIC} (hosting xApps for near-real-time operations)~\cite{Almeida2024}.
These components cooperate to enable flexible, energy-efficient network operations.

The architecture relies on standardized interfaces for closed-loop control:
The \acl{A1}\acused{A1} interface connects the \ac{Non-RT RIC} to the \ac{Near-RT RIC}, enabling policy delivery and configuration updates.
The internal \acl{R1}\acused{R1} interface exposes platform services to rApps.
The \acl{O1}\acused{O1} interface supports management, orchestration, and metrics collection (via the \ac{VES} protocol) from the \ac{RAN} and the \ac{Near-RT RIC} to the \ac{SMO}, which is essential for E2 Node lifecycle management (activation/deactivation).
Within the \ac{Near-RT RIC}, the \ac{RMR} routing bus enables low-latency messaging, while the \ac{SDL} provides a shared data repository.
Finally, the \acl{E2}\acused{E2} interface connects the \ac{Near-RT RIC} to the \ac{RAN} for near-real-time control, utilizing standard service models like \ac{E2SM-KPM} for performance metrics and \ac{E2SM-RC} for control actions.

Energy efficiency is a principal \ac{O-RAN} use case defined by Working Group 1~\cite{O-RAN.WG1.NESUC}.
Supported strategies include Carrier and Cell Switch Off/On, \ac{RF} Channel Reconfiguration, and sleep modes.
Operating in conjunction, the \acp{RIC} and the \ac{SMO} automate these mechanisms based on temporal traffic demands, thereby minimizing network power consumption while preserving coverage and capacity.

%% file: Sections/4-ProblemFormulation.tex
\section{System Model and Problem Formulation}\label{sec:model}

This section formulates the downlink Energy- and \ac{QoS}-aware load balancing problem. We aim to minimize network power consumption by coordinating \ac{O-RU} activation states, transmit-power adjustments, and bandwidth allocations, while satisfying user-association and \ac{QoS} requirements. We focus on the downlink because \ac{O-RU} transmissions dominate RAN energy consumption.

\subsection{System model}
\label{subsec:system_model}

We consider a dense and highly dynamic place of interest, i.e., one in which many users and their communication demands can change over time, such as sports stadiums, event centers, and large public squares. These scenarios require the network to dynamically adapt its resource allocation based on current demand to meet users' \ac{QoS} requirements and reorchestrate user admission through efficient handover operations. We consider a set $\mathcal{T} = \{t_1, t_2, \ldots, t_{|\mathcal{T}|}\}$ of discrete optimization, handover cycles. Within a single cycle $t$, we assume per-UE spatial positions and per-UE throughput demands remain static; across successive cycles, the aggregate active user set, their demands $\{\lambda_{u}^t\}$, and the subset of powered-on O-RUs may change due to ingress/egress and policy-driven activation/deactivation. To model the \ac{RAN} environment, we define a set $\mathcal{U}^t = \{u_1, u_2, \ldots, u_{|\mathcal{U}^t|}\}$ representing the set of users at the place of interest at time step $t \in \mathcal{T}$. Each user $u_i \in \mathcal{U}^t$ is characterized by a throughput demand $\lambda_{u_i}^t \in \mathbb{R}_{\geq 0}$ (bps). Additionally, we consider a set $\mathcal{R} = \{r_1, r_2, \ldots, r_{|\mathcal{R}|}\}$ of O-RUs, where each \ac{O-RU} $r_i \in \mathcal{R}$ has maximum transmission power $\gamma_{r_i} \in \mathbb{R}_{\geq 0}$ and maximum bandwidth $\rho_{r_i} \in \mathbb{R}_{\geq 0}$. It is worth noting that while we formulate the problem in terms of \acp{O-RU} to capture the physical power consumption of radio transmission, in the \ac{O-RAN} architecture described in Section~\ref{sec:architectural-components}, these entities are managed as logical E2 Nodes, i.e., \acp{O-CU} and \acp{O-DU}. For this model, we assume a direct mapping in which each E2 Node controls a single \ac{O-RU}.
Table~\ref{tab:notation} summarizes the notation used throughout this section.

\begin{table}[t]
\centering
\caption{Summary of notation.}
\label{tab:notation}
\resizebox{\linewidth}{!}{%
\begin{tabular}{c|l}
\hline
\cellcolor{DarkBlue!80}\textbf{Symbol} &
\cellcolor{DarkBlue!80}\textbf{Description} \\
\hline
\multicolumn{2}{c}{\cellcolor{Gray!40}\textbf{Sets \& Indices}} \\
\hline
\rowcolor{LightBlue!100}
$\mathcal{T}$ & Set of discrete optimization/handover cycles \\
$t$ & Time step index, $t \in \mathcal{T}$ \\
\rowcolor{LightBlue!100}
$\mathcal{U}^t$ & Set of active \acp{UE} at time step $t$ \\
$u_i$ & \ac{UE} index, $u_i \in \mathcal{U}^t$ \\
\rowcolor{LightBlue!100}
$\mathcal{R}$ & Set of \acp{O-RU} \\
$r_j$ & \ac{O-RU} index, $r_j \in \mathcal{R}$ \\
\hline
\multicolumn{2}{c}{\cellcolor{Gray!40}\textbf{Parameters}} \\
\hline
\rowcolor{LightBlue!100}
$\lambda_{u_i}^t$ & Throughput demand of \ac{UE} $u_i$ at step $t$ (bps) \\
$\gamma_{r_j}$ & Maximum transmission power of \ac{O-RU} $r_j$ \\
\rowcolor{LightBlue!100}
$\rho_{r_j}$ & Maximum bandwidth of \ac{O-RU} $r_j$ \\
$\theta^{RF}_{r_j}$ & Static power consumption of \ac{O-RU} $r_j$ \\
\rowcolor{LightBlue!100}
$\eta_{r_j}$ & Power amplifier efficiency of \ac{O-RU} $r_j$ \\
$\beta(u_i, r_j)$ & Channel gain between \ac{UE} $u_i$ and \ac{O-RU} $r_j$ \\
\rowcolor{LightBlue!100}
$\sigma^2$ & Noise and interference power \\
$S/N$ & Signal-to-Noise Ratio \\
\rowcolor{LightBlue!100}
$\epsilon$ & Small positive constant (minimum active power) \\
\hline
\multicolumn{2}{c}{\cellcolor{Gray!40}\textbf{Decision Variables}} \\
\hline
\rowcolor{LightBlue!100}
$\mathbf{x}^t_{u_i, r_j}$ & Binary: \ac{UE}--\ac{O-RU} association ($\{0,1\}$) \\
$\mathbf{y}^t_{u_i, r_j}$ & Bandwidth allocated to \ac{UE} $u_i$ on \ac{O-RU} $r_j$ \\
\rowcolor{LightBlue!100}
$\mathbf{w}^t_{r_j}$ & Transmission power of \ac{O-RU} $r_j$ \\
$\mathbf{z}^t_{r_j}$ & Binary: \ac{O-RU} activation state ($\{0,1\}$) \\
\hline
\end{tabular}%
}
\end{table}

\subsection{Problem formulation}

We aim to minimize the transmission power of \acp{O-RU} while accommodating all users and meeting their demands. In this case, we define four decision variables to represent this goal: 

\begin{itemize}
    \item $\textbf{x}^t_{u_i, r_j} = \{0, 1\}$, which defines if the user $u_i \in \mathcal{U}^t$ is associated with \ac{O-RU} $r_j \in \mathcal{R}$ at the time step $t \in \mathcal{T}$.
    \item $\textbf{y}^t_{u_i, r_j} \in \mathbb{R}_{\geq 0}$, which defines the bandwidth allocated to the user $u_i \in \mathcal{U}^t$ in O-RU $r_j \in \mathcal{R}$ at the time step $t \in \mathcal{T}$.
    \item $\textbf{w}^{t}_{r_j} \in \mathbb{R}_{\geq 0}$, which represents the transmission power selected for the \ac{O-RU} $r_j \in \mathcal{R}$ at the time step $t \in \mathcal{T}$.
    \item $\textbf{z}^t_{r_j} = \{0, 1\}$, which represents whether the \ac{O-RU} $r_j \in \mathcal{R}$ is activated or not, at the time step $t \in \mathcal{T}$.
\end{itemize}

We formulate our objective function as follows:
\begin{equation}
    	\textit{minimize} \qquad \left ( \sum_{t \in \mathcal{T}}\sum_{r_j \in \mathcal{R}} \frac{\textbf{w}^{t}_{r_j}}{\eta_{r_j}} + \theta^{RF}_{r_j} \right ),
    \label{eq:objective}
\end{equation}

\noindent where $\theta^{RF}_{r_j}$ and $\eta_{r_j}$ represent respectively the static power consumption and the power amplifier efficiency of \ac{O-RU} $r_j \in \mathcal{R}$~\cite{Mai2023,Auer:2011}. This static power consumption is inherent to the \ac{O-RU}'s operation and is independent of dynamic power adjustments based on network demand. We present the problem constraints as follows.

Every user $u_i \in \mathcal{U}^t$ must be associated with one O-RU $r_j \in \mathcal{R}$ at each time step $t \in \mathcal{T}$:
\begin{equation}
    \sum_{r_j \in \mathcal{R}} \textbf{x}^t_{u_i, r_j} = 1, \qquad \forall u_i \in \mathcal{U}^t, t \in \mathcal{T}.
    \label{eq:association}
\end{equation}

The transmission power $\textbf{w}^{t}_{r_j}$ assigned to \ac{O-RU} $r_j \in \mathcal{R}$ at time step $t \in \mathcal{T}$ must be positive and respect its maximum power capacity $\gamma_{r_j}$:
\begin{equation}
    0 \leq \textbf{w}^{t}_{r_j} \leq \gamma_{r_j}, \qquad \forall r_j \in \mathcal{R}, t \in \mathcal{T}.
    \label{eq:power-cap}
\end{equation}

If the \ac{O-RU} $r_j \in \mathcal{R}$ is inactive ($\textbf{z}^t_{r_j} = 0$) at time step $t \in \mathcal{T}$, its transmission power $\textbf{w}^{t}_{r_j}$ must be zero:
\begin{equation}
    	\textbf{w}^{t}_{r_j} \leq \textbf{z}^t_{r_j} \gamma_{r_j}, \qquad \forall r_j \in \mathcal{R}, t \in \mathcal{T}.
    \label{eq:inactive-power-link}
\end{equation}

If the \ac{O-RU} $r_j \in \mathcal{R}$ is inactive ($\textbf{z}^t_{r_j} = 0$) at time step $t \in \mathcal{T}$,  then there must be no users associated with $r_j \in \mathcal{R}$:
\begin{equation}
    \sum_{u_i \in \mathcal{U}^t}\textbf{x}^{t}_{u_i,r_j} \leq \textbf{z}^t_{r_j} |\mathcal{U}^t|, \qquad \forall r_j \in \mathcal{R}, t \in \mathcal{T}.
    \label{eq:inactive-assoc}
\end{equation}

If the \ac{O-RU} $r_j \in \mathcal{R}$ is active ($\textbf{z}^t_{r_j} = 1$), its transmission power $\textbf{w}^{t}_{r_j}$ must be positive and greater than zero:
\begin{equation}
        \textbf{w}^{t}_{r_j} \geq \textbf{z}^t_{r_j} \epsilon, \qquad \forall r_j \in \mathcal{R}, t \in \mathcal{T},
    \label{eq:active-min-power}
\end{equation}

\noindent where $\epsilon$ is a small positive constant.

The total bandwidth $\sum_{u_i \in \mathcal{U}^t} \textbf{y}^t_{u_i, r_j}$ used by an \ac{O-RU} $r_j \in \mathcal{R}$ that is active ($\textbf{z}^t_{r_j} = 1$) must not exceed its maximum bandwidth capacity $\rho_{r_j}$, this can be expressed as:
\begin{equation}
    \sum_{u_i \in \mathcal{U}^t} \textbf{y}^t_{u_i, r_j} \leq \rho_{r_j} \textbf{z}^t_{r_j} \qquad \forall r_j \in \mathcal{R}, t \in \mathcal{T}.
    \label{eq:bandwidth-cap}
\end{equation}

To ensure that the allocated bandwidth $\textbf{y}^t_{u_i, r_j}$ meets the throughput demand $\lambda_{u_i}^t$ of every user $u_i \in \mathcal{U}^t$ at each time step $t$, we employed the Shannon's capacity equation to design the last constraint:
\begin{equation}
    \sum_{r_j \in \mathcal{R}} \left ( \textbf{y}^t_{u_i, r_j} \cdot \log \left (1 + \frac{S}{N} \right ) \right ) \geq \lambda_{u_i}^t \qquad \forall u_i \in \mathcal{U}^t, t \in \mathcal{T},
    \label{eq:throughput}
\end{equation}
where $\frac{S}{N}$ is the \ac{SNR}, computed for a user $u_i \in \mathcal{U}^t$ associated with an \ac{O-RU} $r_j \in \mathcal{R}$ using the channel gain $\beta(u_i, r_j)$ (incorporating path loss and shadowing) and thermal noise power $\sigma^2$:
\begin{equation}
    \frac{S}{N} = \frac{\beta(u_i, r_j) \textbf{w}^{t}_{r_j}}{\sigma^2}.
    \label{eq:snr}
\end{equation}

The proposed formulation (as written with the Shannon-based throughput constraint) is a mixed-integer non-linear program (MINLP) and therefore finding its optimal solution is NP-hard. The complexity of MINLP models grows with the number of decision variables and constraints. In our formulation, both the number of decision variables and constraints are dominated by the number of users ($|\mathcal{U}|$), since the number of users is much higher than the number of \acp{O-RU} in practical \ac{vRAN} networks. Solving large-scale instances to proven optimality is computationally expensive. Consequently, we introduce a heuristic approach to obtain high-quality feasible solutions rapidly for practical scenarios with many \acp{O-RU} and users. For benchmarking on small instances, we also employ a linearized MILP surrogate of the original MINLP that is solvable by IBM CPLEX.

\subsection{Heuristic solution}

In this section, we present EQLite, a heuristic algorithm designed to solve the proposed problem formulation with a focus on scalability in practical deployment scenarios. EQLite is a deterministic, evolutive greedy heuristic that begins with all \acp{O-RU} powered off and iteratively activates them in response to variations in user demand. Furthermore, EQLite features an adaptive bandwidth allocation mechanism that dynamically adjusts resource distribution based on the users' wireless channel conditions and traffic demand fluctuations.

\begin{algorithm}[!ht]
\footnotesize
    \caption{Energy- and QoS-aware Load-balancing ITErative heuristic (EQLite)}
    \label{alg:heuristic}
    \KwIn{$t \in \mathcal{T}, \mathcal{U}^t, \lambda^t_{u}$ and $\mathcal{R}$} 
    \KwOut{User association, O-RU transmission power selection and users bandwidth allocation}
    $R \gets \emptyset;$ $A \gets \emptyset;$ \\
    \For{$r \in \mathcal{R}$}{
        $P_{user}^r \gets$ list of users sorted by SNR \\
        $BW_r \gets \rho_r;$ \\
        \For{$u \in \mathcal{U}^t$}{
            $Association[r][u] \gets \textbf{False};$ \\
            $Allocation[r][u] \gets 0;$ \\
        }
    }
    \While{\textbf{True}}{
        $b \gets$ ActivateRU($P_{user}^r$)$;$ \\
        $R$.push($b$)$;$ $b^{TP} \gets \gamma_b;$ \\
        \For{$u \in P_{user}^b$}{
            \If{$u \notin A$}{
                \If{$BW_r \geq \lambda_u^t$}{
                    $Association[b][u] \gets \textbf{True};$ \\
                    $A$.push($u$)$;$ \\
                    $Allocation[r][u] \gets \lambda_u^t;$ \\
                    $BW_r \gets BW_r - \lambda_u^t;$
                }
            }
            \If{$|A| = |\mathcal{U}^t|$ \textbf{or} $|R| = |\mathcal{R}|$}{
                \textbf{break};
            }
        }
    }
    \For{$r \in R$}{
        \While{\textbf{True}}{
            $feasible \gets DecreaseTP(r, Association[r]);$ \tcp{return True or False}
            \If{\textbf{not} $feasible$}{
                \textbf{break};
            }
        }
    }
\end{algorithm}

We present the EQLite heuristic in Algorithm~\ref{alg:heuristic}. Initially, in line 1, the algorithm creates two empty lists to store the set of activated \acp{O-RU} and the set of admitted users, respectively. Moreover, two matrices are initialized to store user associations and bandwidth allocations between users and \acp{O-RU}. From lines 8 to 19, the core logic of the heuristic is executed iteratively. In each iteration, the heuristic activates a new \ac{O-RU} based on users' \ac{SNR}, prioritizing the \ac{O-RU} serving users with the highest link quality. Once selected, the \ac{O-RU} is activated and assigned its maximum transmission power as a baseline configuration. Next, from lines 11 to 17, the algorithm iterates over all users and attempts to associate each with the newly activated \ac{O-RU}. If the \ac{O-RU} has sufficient capacity, the user is admitted and resources are allocated accordingly. If not, the user is denied and the algorithm proceeds to activate another \ac{O-RU} in the next iteration. At the end of the algorithm, the list of activated \acp{O-RU} and associated users is finalized. A final power minimization step is performed to iteratively reduce the transmission power of each active \ac{O-RU}, provided that users' \ac{QoS} requirements remain satisfied.

The computational complexity of the heuristic is expressed as $\mathcal{O}(|\mathcal{T}||\mathcal{R}||\mathcal{U}^t| + |R||A|)$. However, since $R \subset \mathcal{R}$ and $A \subset \mathcal{U}^t$, the complexity can be approximated as $\mathcal{O}(|\mathcal{R}||\mathcal{U}^t|)$. Since in practice, \ac{vRAN} networks present a higher number of users than \acp{O-RU}, the complexity of our heuristic algorithm is dominated by $\mathcal{O}(|\mathcal{U}^t|)$. Having established the system model and formulated the optimization problem for energy-efficient resource allocation in \ac{O-RAN} networks, we next detail the architecture of our proposed solution. In the following section, we describe how we implement this optimization using \ac{O-RAN}'s rApps and xApps, and how the various architectural components interact to enable energy savings.

\subsection{Analytical evaluation}
\label{subsec:analytical evaluation}

We compare the quality of the solutions obtained from three approaches to evaluate our proposed problem formulation: (i) the optimization model solved using CPLEX, (ii) our heuristic algorithm implemented in Python, and (iii) a baseline algorithm that associates users with the \ac{O-RU} offering the highest \ac{SINR} level. The baseline reflects the behavior typically observed in current real-world cellular networks. We consider two evaluation scenarios because the optimization model has limited scalability compared to non-exact approaches (heuristic and baseline). In the first scenario, we use a \ac{vRAN} topology with 12 \acp{O-RU} and up to 1,024 simultaneous users. In the second scenario, we consider a topology with 56 \acp{O-RU} and up to 10,240 users.

\begin{figure}
    \centering
    \includegraphics[width=1\linewidth]{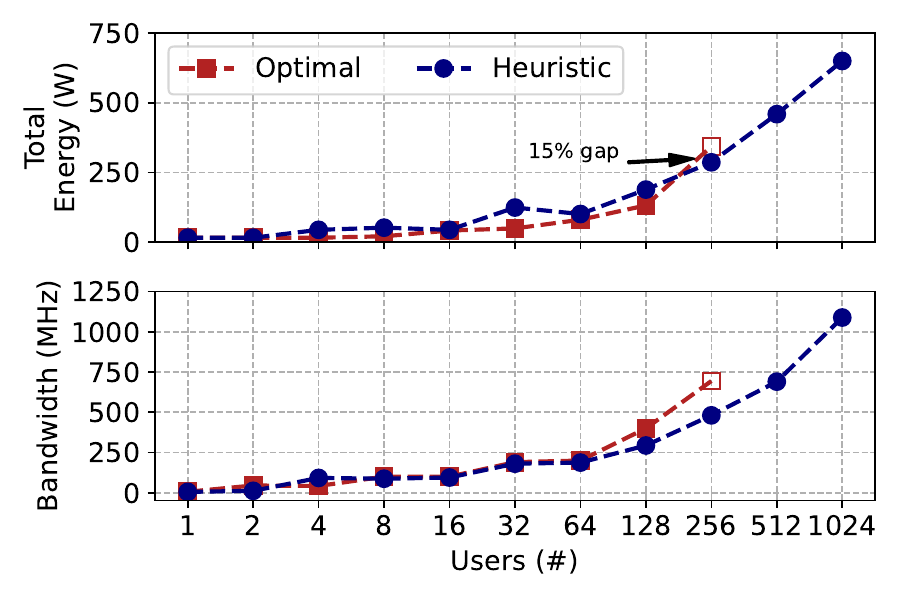}
    \caption{Comparing optimal and heuristic solutions.}
    \label{fig:optimal_heuristic}
\end{figure}

Fig.~\ref{fig:optimal_heuristic} shows the comparison between the optimal and heuristic solutions. The optimization model consistently achieves the lowest total energy consumption, considering \acp{O-RU} transmission power and static energy consumption, across all scenarios. Nevertheless, our heuristic yields results close to optimal, even matching the optimization model in scenarios with fewer users. Regarding bandwidth allocation, the heuristic tends to use fewer bandwidth resources than the optimal model, offsetting its slightly higher energy consumption while still meeting users' \ac{QoS} requirements.

As the number of users increases, competition for resources intensifies, and the optimality gap of the heuristic grows, reaching up to 40\% in the worst-case scenario in our experiments. However, the optimization model itself fails to scale, and it cannot provide solutions in a timely manner for larger scenarios. With a 24-hour time limit, the model achieves only a best-integer solution for 256 users, with a 15\% gap to the best bound. For scenarios with 512 and 1,024 users, the optimization model does not find any feasible solution within this time interval.

\begin{figure}
    \centering
    \includegraphics[width=1\linewidth]{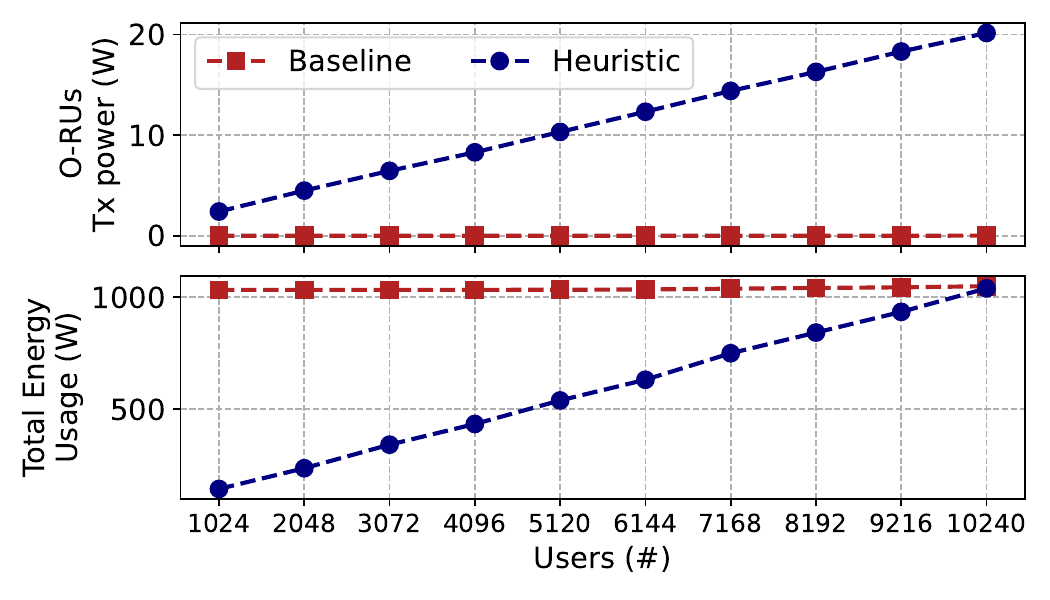}
    \caption{Comparing heuristic and baseline solutions.}
    \label{fig:heuristic_baseline}
\end{figure}

Fig.~\ref{fig:heuristic_baseline} compares our heuristic algorithm with the baseline for scenarios with a larger number of users. The baseline algorithm consistently achieves lower \ac{O-RU} transmission power because users experience higher \ac{SINR}. In contrast, our heuristic solution results in a total transmission power of 20 W across the 52 \acp{O-RU}, corresponding to an average of 0.39 W per \ac{O-RU}. However, when analyzing the total energy consumption of \acp{O-RU}, our heuristic achieves considerably lower values than the baseline. This improvement arises from the fact that, on average, 44\% of \acp{O-RU} remain switched off in our solution. By deactivating underutilized \acp{O-RU}, the heuristic achieves higher energy efficiency than the baseline, even though it operates with higher transmission power concentrated on a smaller number of active \acp{O-RU}.

\begin{figure}
    \centering
    \includegraphics[width=.8\linewidth]{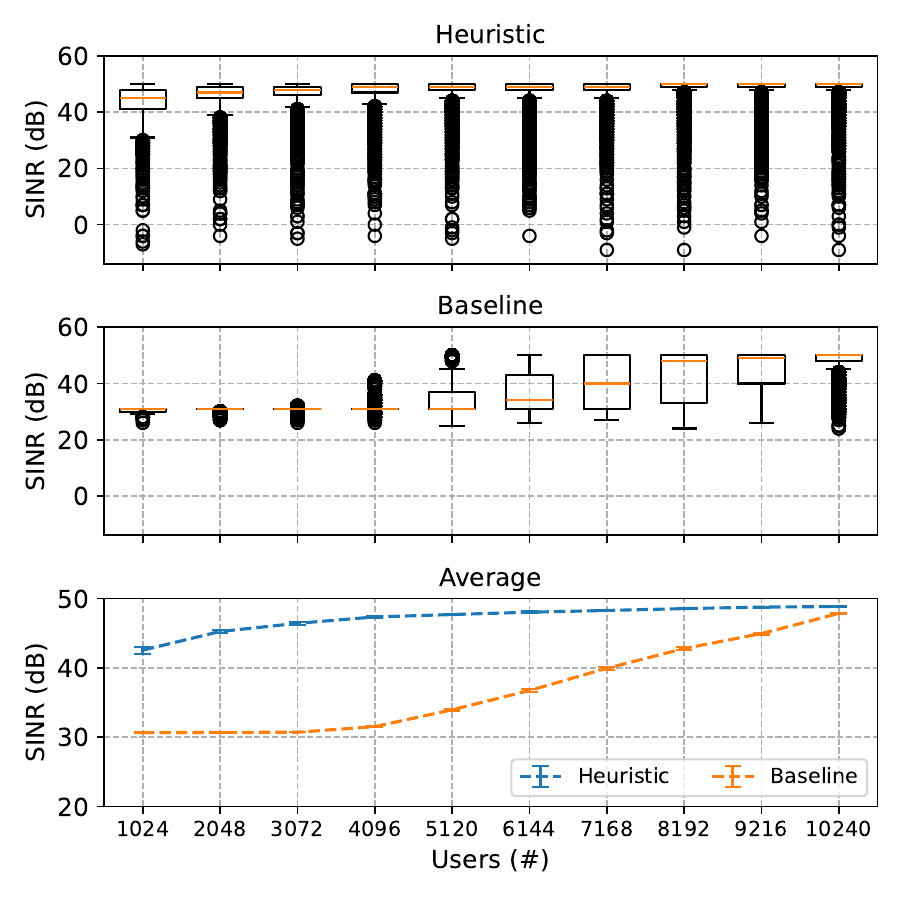}
    \caption{Comparing SINR for heuristic and baseline solutions.}
    \label{fig:SINR_comparison}
\end{figure}

Fig.~\ref{fig:SINR_comparison} shows the \ac{SINR} distribution and the average \ac{SINR} in the second scenario. Although our proposed heuristic produces more low-\ac{SINR} outliers than the baseline, it achieves a higher average \ac{SINR}. This behavior stems from its more effective load balancing across active \acp{O-RU}, which better aligns user demand with available resources. In contrast, the baseline relies on a greedy user-association strategy, resulting in suboptimal distribution of users among the active \acp{O-RU}. Consequently, while some users under the heuristic experience lower \ac{SINR}, the overall network operates more efficiently, leading to an improved average \ac{SINR}. These results indicate that our proposed heuristic enhances resource utilization and energy efficiency while improving the overall wireless link quality between users and \acp{O-RU}.

\begin{figure}
    \centering
    \includegraphics[width=0.7\linewidth]{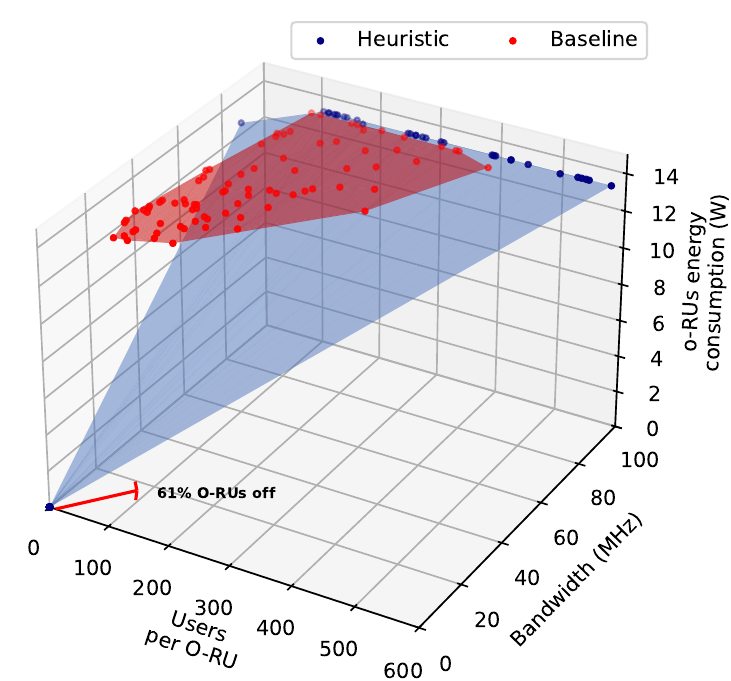}
    \caption{Overall comparison of heuristic and baseline solutions.}
    \label{fig:3D_plot}
\end{figure}

Fig.~\ref{fig:3D_plot} presents an overall comparison between our heuristic and the baseline model for the scenario with 4,096 users, providing a zoomed-in view of the results. While the baseline model shows energy consumption per \ac{O-RU} values close to those of the heuristic, it keeps all \acp{O-RU} powered on, which leads to higher total energy usage, as previously discussed. In contrast, our heuristic activates only 39\% of the available \acp{O-RU} to cover and guarantee \ac{QoS} for all 4,096 users, leaving 61\% of \acp{O-RU} in standby mode. This strategy increases the number of users served per active \ac{O-RU} and consequently raises the bandwidth utilization of the active \acp{O-RU}, but it still results in significantly lower total energy consumption.

\subsection{Optimization Intervals}
Executing the optimization procedure every 5--15 minutes strikes a balance between reacting to load dynamics and avoiding excessive reconfiguration overhead~\cite{Zanzi2020, Baena2021}. This periodic Non-RT planning loop maps directly to the Non-RT RIC timescale ($>1$\,s), while the Near-RT RIC is responsible solely for executing the handover actions.

%% file: Sections/5-Architecture.tex
\section{Architecture for Energy Savings in O-RAN Networks} \label{sec:architectural-components}

\begin{figure}[t]
\centering
\includegraphics[width=.8\linewidth]{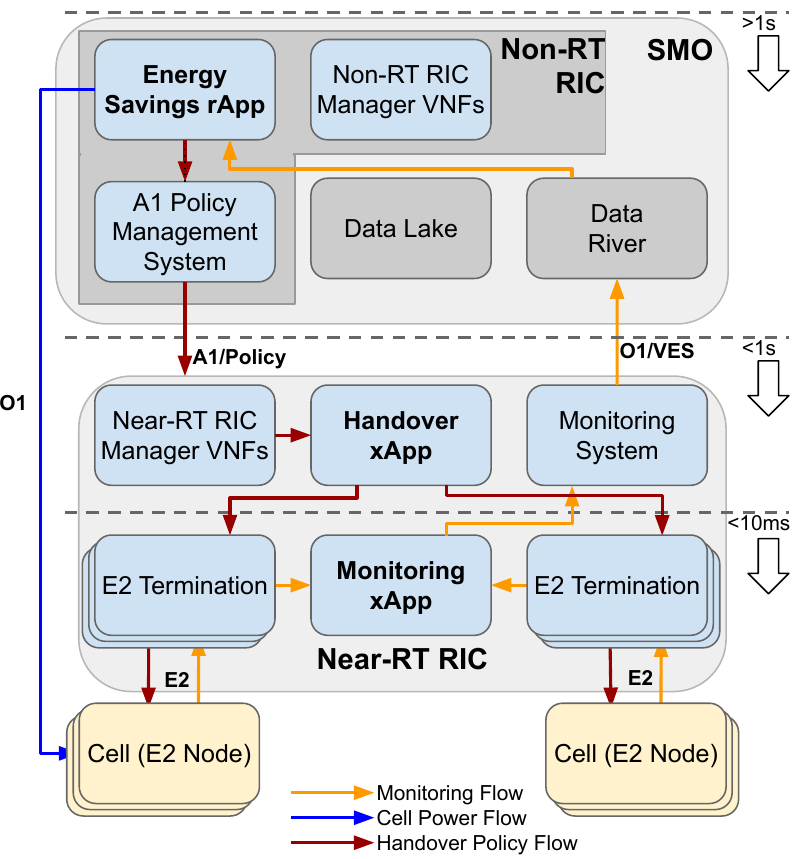}
\caption{Architecture of our proposed energy saving O-RAN solution.}
\label{fig:architecture}
\end{figure}

To address the increasing need for energy efficiency in mobile networks, we propose an \ac{O-RAN}-based control architecture that embeds the proposed energy- and \ac{QoS}-aware load-balancing optimization model into \acp{RIC}.
The \ac{Non-RT RIC} and \ac{SMO} host an Energy Savings rApp that periodically solves the mixed-integer optimization problem governing \ac{UE} reassociation, E2 Node activation/deactivation, and transmit-power selection.
The resulting high-level policies are forwarded to the \ac{Near-RT RIC}, where a Handover xApp enforces them on the \ac{E2} interface via near real-time mobility actions.
This design constitutes a hierarchical two-tier control architecture: the Energy Savings rApp operates as a periodic planning loop within the \ac{Non-RT RIC} timescale ($>$1\,s, typically minutes), while the Handover xApp enforces per-\ac{UE} decisions within the \ac{Near-RT RIC} timescale (10\,ms--1\,s).
Table~\ref{tab:timescale-mapping} summarizes this mapping.
By jointly exploiting non real-time and near real-time management functions, the solution coordinates \ac{UE} handovers, E2 Node lifecycle management, and resource allocation across E2 Nodes to reduce network energy consumption while preserving the \ac{QoS} guarantees of all \acp{UE}.
Fig.~\ref{fig:architecture} illustrates the overall architecture and the interactions among the \ac{Non-RT RIC}, the \ac{SMO}, and the \ac{Near-RT RIC}.
The remainder of this section details the control architecture, the interfaces and data flow, and the implementation of the energy savings and handover mechanisms.

\subsection{End-to-end Control Architecture}

In the proposed end-to-end architecture, the Energy Savings rApp in the \ac{Non-RT RIC} leverages historical data stored in the \ac{SMO}, e.g., within the Data Lake and Data River, and \acp{KPI} aggregated over time intervals of 5 to 15 minutes to decide when E2 Nodes are activated, deactivated, or placed into low-power states. These non real-time decisions define high-level energy-saving policies and target load levels for each E2 Node.

The Handover xApp in the \ac{Near-RT RIC} enforces these policies at finer time scales.
It responds to short-term dynamics, including sudden traffic surges and rapid channel-quality variations, by issuing handover commands on a millisecond-to-second timescale to redistribute \acp{UE} and preserve their \ac{QoS}.
By assigning coarse-grained energy and load objectives to the Energy Savings rApp and fine-grained mobility control within those objectives to the Handover xApp, the architecture avoids conflicting decisions and ensures that near real-time actions remain consistent with non real-time energy-efficiency targets.
Because the Handover xApp executes only rApp-directed policies and does not independently trigger cell (de)activation, cell-state changes occur exclusively at \ac{Non-RT RIC} cadence ($\sim$535\,s per cycle), which provides implicit hysteresis against ON/OFF oscillation (ping-pong).

\begin{table}[t]
\centering
\caption{Timescale mapping of the hierarchical control architecture.}
\label{tab:timescale-mapping}
\resizebox{\linewidth}{!}{%
\begin{tabular}{c|c|c|c|c}
\hline
\cellcolor{DarkBlue!80}\textbf{\shortstack{O-RAN\\Tier}} &
\cellcolor{DarkBlue!80}\textbf{Component} &
\cellcolor{DarkBlue!80}\textbf{\shortstack{O-RAN\\Timescale}} &
\cellcolor{DarkBlue!80}\textbf{\shortstack{Measured\\Cycle}} &
\cellcolor{DarkBlue!80}\textbf{Role} \\
\hline
\rowcolor{LightBlue!100}
\shortstack{Non-RT\\RIC} & \shortstack{Energy Savings\\rApp} & $>$1\,s & $\sim$535\,s & \shortstack{Periodic policy computation:\\cell (de)activation, Tx power,\\UE reassociation} \\
\shortstack{Near-RT\\RIC} & \shortstack{Handover\\xApp} & \shortstack{10\,ms\\--1\,s} & \shortstack{0.12--0.20\\\,s/UE} & \shortstack{Per-UE handover enforcement\\via E2 interface} \\
\hline
\end{tabular}%
}
\end{table}

\subsection{Interfaces and Data Flow}

Three primary interfaces support communication between the components of the architecture: the \ac{O1}, \ac{A1}, and \ac{E2} interfaces. The \ac{O1} interface is responsible for E2 Node lifecycle management, i.e., turning E2 Nodes on or off, and collecting their performance metrics. The \ac{A1} interface transfers high-level policies specifying which E2 Nodes should be deactivated or kept active, along with target load levels and desired \ac{UE} assignments. Finally, the \ac{E2} interface provides near real-time control and monitoring of E2 Nodes, enabling the Handover xApp to issue handover commands.

\subsection{Implementation of Energy Savings and Handover}

The Energy Savings rApp and Handover xApp coordinate via continuous telemetry and a standard message flow. The Monitoring xApp continuously exports aggregated radio and metric updates to the rApp. When a reconfiguration is triggered, the rApp issues O1 activation/deactivation commands to the E2 Nodes and transmits a prescriptive UE-migration policy via A1 to the Handover xApp, which enforces E2AP-based handovers to transition the target UEs.

\subsubsection*{Energy-Policy Schema}

Each entry defines an E2 Node using its \ac{MCC}, \ac{MNC}, NodeB identifier (field: nodebId), and \ac{PCI}, along with the list of \acp{UE} that it is currently serving (field: ueList, entries identified by \acp{IMSI}), following the E2 Node–UE energy-policy schema presented in Listing~\ref{lst:e2node-schema}. This mapping enables our rApp to select nodes for power reduction or deactivation and provides our xApp with the exact list of \acp{UE} that need to be migrated to neighboring E2 Nodes, ensuring their service continuity without disruption.

\begin{lstlisting}[language=json,
  caption={JSON-Schema of the E2NodeList element.},
  label={lst:e2node-schema},
  linewidth=\columnwidth,
  frame=single,
  numbers=left,
  numberstyle=\tiny\color{gray},
  breaklines=true,
  keepspaces=true,
  basicstyle=\ttfamily\scriptsize]
{ "E2NodeList": {
    "type": "array",
    "items": {
      "type": "object",
      "properties": {
        "mcc":     { "type": "string" },
        "mnc":     { "type": "string" },
        "nodebId": { "type": "string" },
        "pci":     { "type": "string" },
        "ueList": {
          "type": "array",
          "items": {
            "type": "object",
            "properties": { "imsi": { "type": "string" } }
          } } } } } }
\end{lstlisting}

The \ac{A1} framework defined in the \ac{O-RAN} \ac{A1} Application Protocol~\cite{O-RAN.WG2.A1AP} supports custom policy type registration via \ac{JSON} Schema, enabling use-case-specific types when the standardized \ac{A1} Type Definitions~\cite{O-RAN.WG2.A1TD} do not yet cover a particular operational need.
Our energy-policy schema exploits this extensibility mechanism for three reasons.
First, the standardized \ac{A1} Traffic Steering Preference (TSP) policy is declarative: it expresses per-cell preferences (e.g., PREFER/FORBID) and delegates enforcement logistics to the \ac{Near-RT RIC}.
Our use case requires a prescriptive policy that specifies the exact set of \ac{E2} Nodes to (de)activate together with the per-\ac{UE} handover list and target-cell assignments, information that TSP's scope model does not carry.
Second, the schema bundles cell-lifecycle and \ac{UE}-migration semantics into a single atomic message, ensuring that the Handover xApp receives a self-contained action plan and avoiding race conditions between independent \ac{A1} and \ac{O1} operations.
Third, at the time of prototyping (\ac{OSC} I/J releases), the \ac{A1} Mediator implemented only A1-AP v1, which predated the A1-TD type definitions published in later \ac{O-RAN} ALLIANCE specification releases; registering a custom policy type was the documented integration path.
Migrating to a standardized A1-TD energy-saving policy type, once one with per-\ac{UE} reassociation semantics is ratified, is an incremental engineering step that does not alter the architecture or the control flow presented above.

%% file: Sections/6-Prototype.tex
\section{Prototype} \label{sec:prototype}

In this section, we demonstrate our solution in a practical environment, which led us to implement certain missing components within the \ac{OSC} platform. These implementations are essential to the feasibility of our Energy Saving solution and constitute a significant part of our contribution. We detail the roles of these components and illustrate how rApps and xApps interact within the \ac{O-RAN} architecture to achieve energy-efficient network operations.

\subsection{Enhancements over the OSC}

We significantly modified several components within the \ac{OSC} platform to overcome the challenges identified in Section~\ref{sec:architectural-components}. These enhancements are essential for the feasibility of the Energy Savings solution\footnote{The software artifacts and implementation details are available at: \url{https://github.com/zanattabruno/Energy-Saver-Tests}}.
Firstly, we enhanced the \ac{VESPA} Manager in the \ac{Near-RT RIC} and the \ac{VES} collector in the \ac{SMO} to efficiently integrate and manage event streams. The default configurations provided by the \ac{OSC} were inadequate for our requirements. Modifications include increased flexibility for dynamic configuration changes and enhanced data handling capacity. For instance, the \ac{VES} collector URL is now configurable via ConfigMap files, and the maximum HTTP request size has been increased to accommodate larger data volumes. These enhancements enable real-time processing of performance metrics and events, which is critical for dynamic resource management and energy saving operations.
Secondly, we modified the A1 Mediator on the \ac{Near-RT RIC}. The current \ac{Non-RT RIC} release supports only version 1 of the A1 \ac{API}, which contains a bug that causes policy instances to be sent without the subId. Our modifications ensure policy instances include the necessary ID for proper functionality and integration, facilitating seamless communication between the \ac{Non-RT RIC} and \ac{Near-RT RIC}.
Thirdly, we addressed persistent storage in \ac{K8s}. In this context, K8s lacks a default storage class that enables pods to use persistent storage. We implemented the Local Path Provisioner to overcome this limitation, allowing pods to request storage resources dynamically. This issue is crucial for maintaining stateful information and ensuring data persistence across system reboots or crashes.

We also extended the \texttt{E2sim} tool to simulate \ac{UE}, radio metrics, and handover procedures through a \ac{REST} \ac{API}, providing a robust emulation environment for evaluating our energy- and QoS-aware load-balancing approach under realistic conditions. In this case, we developed the \ac{RFSS}, a modular simulator that models radio units, grandstand tiers, and \acp{UE}, and manages \ac{O-FH} signaling and data exchange to instantiate and control \acp{UE} within our experimental \ac{RAN} setup. This tool computes \ac{UE} metrics, e.g., \ac{RSRP}, \ac{RSRQ}, \ac{CQI}, \ac{SINR}, and \ac{BLER}, for delivery to \texttt{E2sim} and simulates handover events. Furthermore, we implemented a Monitoring xApp designed around the RAN-centric control loop. It establishes \ac{RAN} Control subscriptions, decodes the resulting \ac{RIC} Indications, and exports the extracted \ac{UE} and cell \acp{KPI} through a Prometheus metrics sink.

\subsection{Energy Savings rApp} \label{subsubsec:rapp-energy-optimizer}

The Energy Savings rApp operates within the \ac{Non-RT RIC} and executes the optimization algorithm to minimize energy consumption subject to user-demand constraints.
It processes historical and periodically aggregated data on network performance, user mobility patterns, and energy usage.
The underlying NP-hard optimization problem is originally a MINLP, i.e., for small-instance ground-truth and periodic calibration.
We use a linearized MILP surrogate solved with IBM CPLEX 22.1.0 (branch-and-bound and cutting planes).
In the deployed periodic Non-RT control loop, we rely on the lightweight \ac{EQLite} heuristic to generate policies within strict latency budgets.
These policies, which specify which E2 Nodes should be active or in standby mode, are communicated to the \ac{Near-RT RIC} through the \ac{A1} interface.

\subsection{Handover xApp} \label{subsec:xapp-energy-saver}

The Handover xApp operates within the \ac{Near-RT RIC} and implements the policies generated by the Energy Savings rApp.
It receives policies via the \ac{A1} interface and determines the required actions.
It executes handover procedures to transition \acp{UE} from E2 Nodes scheduled for deactivation to active E2 Nodes, and continuously monitors the network to respond promptly to changes in user demand or network conditions.
The handover process is carefully managed to ensure seamless transitions for \acp{UE}.
The Handover xApp uses specific \ac{E2SM-RC} messages to instruct E2 Nodes to initiate handovers, monitor progress, and confirm successful completion before the rApp deactivates the E2 Node via the \ac{O1} interface.

\subsection{Components Interaction Overview}

Our prototype employs a dynamic load balancing mechanism that continuously monitors user demand to trigger E2 Node on/off states over time.
Within each optimization, handover cycle, we assume static \ac{UE} spatial positions and per-\ac{UE} throughput requirements (no intra-cycle mobility).
In this case, adaptation occurs between cycles as the active \ac{UE} set and powered-on E2 Nodes evolve.
Whenever the rApp detects a sustained decrease in the number of \acp{UE} or in aggregate throughput (with thresholds of 20\% for both), it issues updated energy saving policies specifying which E2 Nodes can be placed in low-power mode.
The Handover xApp hands off any remaining \acp{UE} in those E2 Nodes' coverage area to neighboring E2 Nodes with sufficient capacity.
Conversely, if traffic demand surges in a previously dormant cell region, the rApp activates the E2 Node and sends a new policy via the \ac{A1} interface.
Moreover, the Handover xApp ensures that \acp{UE} are promptly re-associated to maintain the desired \ac{QoS}.
Over time, this cycle of deactivating underutilized E2 Nodes and reactivating them as demand increases helps minimize energy consumption while ensuring a satisfactory user experience.

\begin{figure*}[t]
\centering
\includegraphics[width=.65\linewidth]{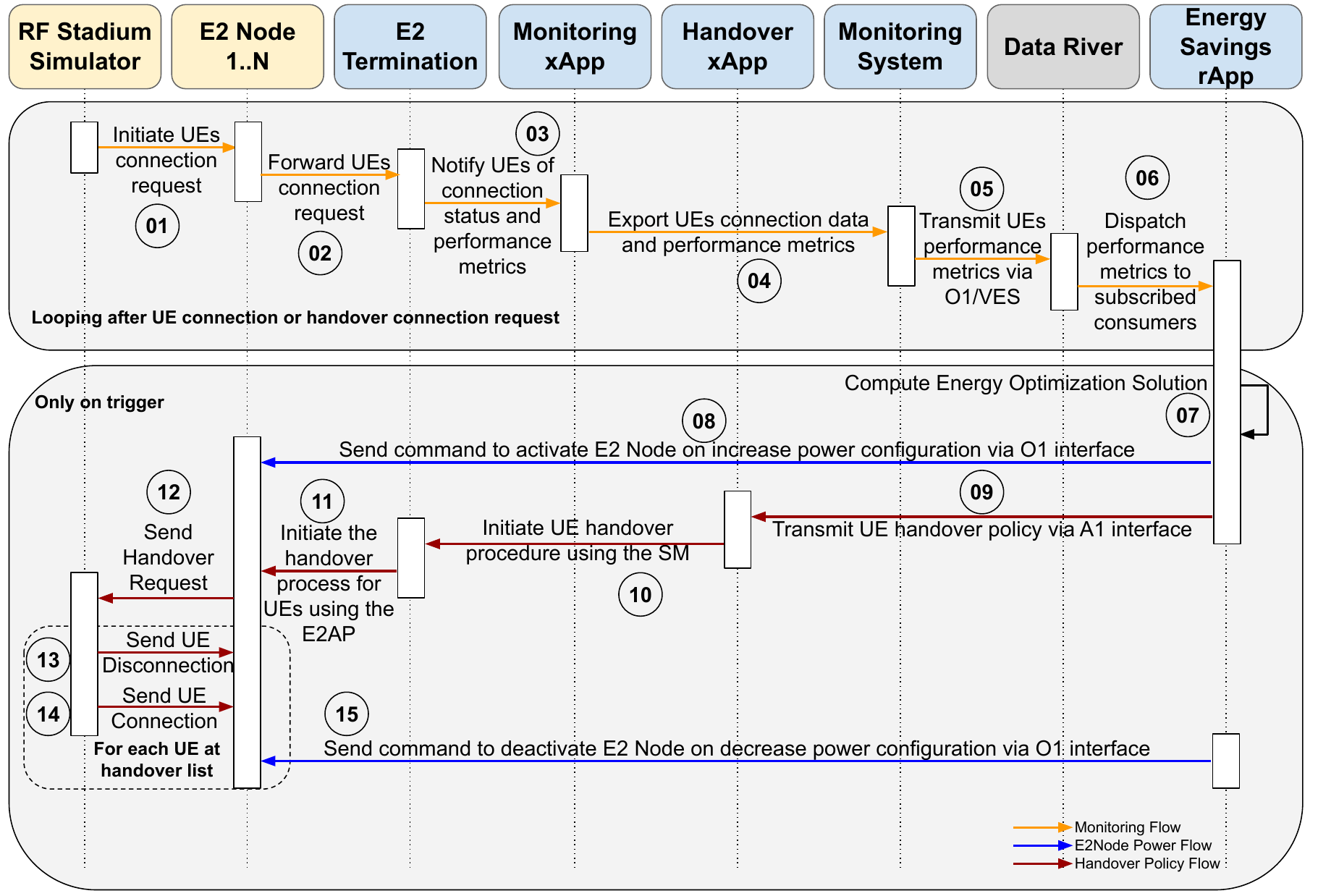}
\caption{Sequence of interactions for the energy saving use case.}
\label{fig:Prototype-Sequence}
\end{figure*}

Fig.~\ref{fig:Prototype-Sequence} illustrates the two-part workflow for our Energy Saving use case as a single continuous loop, with additional steps triggered as needed. In the upper portion (Steps 1–6), the \ac{RFSS} initiates a \ac{UE} connection request (Step 1), which the E2 Node forwards (Step 2); the Monitoring xApp notifies the system of the new \ac{UE} connection and performance metrics (Step 3), exports these data (Step 4), and transmits them to the \ac{SMO} via O1/\ac{VES} (Step 5). The \ac{SMO}’s Data River relays these metrics to subscribed consumers, including the Energy Savings rApp (Step 6). Once specific triggers occur (lower portion, Steps 7–15), the rApp computes an energy savings solution (Step 7), possibly sending a command to increase an E2 Node’s power configuration (Step 8) and updated handover policies via the \ac{A1} interface (Step 9). The Handover xApp initiates the \ac{UE}‐handover procedure (Step 10) via \ac{E2AP} (Step 11), triggering handover requests (Step 12) and subsequent \ac{UE} disconnections and reconnections (Steps 13–14). Finally, once traffic has been shifted, the rApp may reduce power settings by issuing a command to deactivate the E2 Node (Step 15), balancing service quality and energy savings.

By distributing the optimization tasks between the \ac{Non-RT RIC} and \ac{Near-RT RIC} and enhancing existing \ac{O-RAN} components, we demonstrate how the \ac{O-RAN} architecture can enable energy-efficient solutions in the \ac{RAN}. Our use case illustrates the practical implementation of these concepts, showcasing the potential for significant energy savings in dynamic network environments. Because the \ac{RAN} side is emulated by the \ac{RFSS} and \ac{E2Sim}, the prototype validates control-plane behavior and software integration over production \ac{O-RAN} interfaces rather than RF hardware performance; Section~\ref{sec:energy-results} delineates the exact experimental scope. Building upon this prototype, we conduct experimental evaluations to assess the performance of our energy saving approach, as discussed in the next section.

%% file: Sections/7-Evaluation.tex
\section{Experimental Evaluation} \label{sec:energy-results}

We validate the \ac{O-RAN} architecture as a platform for developing and evaluating energy-efficient solutions for future 5G Advanced networks. We first describe our proof-of-concept implementation of the energy-efficient solution proposed in Section~\ref{sec:model} within a practical \ac{O-RAN} environment that manages a simulated \ac{RAN}. The Energy Savings rApp used in these experiments relies on the implementation of the \ac{EQLite} heuristic. Moreover, we assess the performance of our solution with respect to energy savings, handover delay, end-to-end control delay, computational footprint, and throughput feasibility. Throughout the evaluation, we adopt a discrete-time closed-loop viewpoint: within each optimization, handover cycle, the set of \ac{UE} positions and their per-\ac{UE} throughput requirements are fixed (no intra-cycle mobility), while between cycles users may arrive/depart (changing $|\mathcal{U}^t|$ and demand mix). \acp{O-RU} may be (de)activated according to the energy saving policy. This periodic, discrete-time approach is consistent with the \ac{Non-RT RIC} operational model defined by the O-RAN ALLIANCE, where energy-management decisions are computed on a timescale of minutes and enforced by the \ac{Near-RT RIC} on a sub-second timescale (see Table~\ref{tab:timescale-mapping}). This approach isolates the impact of load evolution and \ac{O-RU} state adaptation. Our throughput analysis reveals a trade-off between energy efficiency and strict demand fulfillment: aggressive cell deactivation at low loads yields roughly 50\% \ac{QoS} satisfaction, which degrades further at higher densities (Section~\ref{subsec:qos-analysis}). The evaluation directly measures the objective in~\eqref{eq:objective} (energy/power) while respecting the association constraint and bandwidth-capacity constraint described in Section~\ref{sec:model}.

Before presenting the results, we delineate the experimental scope.
On the \ac{RAN} side, \ac{E2} Nodes are emulated by \ac{RFSS} and \ac{E2Sim}: the signaling plane is exercised over production \ac{O-RAN} interfaces, but PHY-layer processing and real RF propagation are abstracted.
On the observability side, experiments up to 1{,}024~\acp{UE} exercise the full \ac{SMO} data-path (Monitoring xApp, \ac{VESPA}, ONAP \ac{VES}, Kafka).
Beyond 1{,}024~\acp{UE}, payload limits and serialization overheads (see Section~\ref{subsec:e2e-delay}) require bypassing the \ac{VES} layer to feed metrics directly from Prometheus to the rApp.
Consequently, the testbed validates functional correctness and control-loop integration at all scales, and full observability-pipeline scalability up to 1{,}024~\acp{UE}.

\subsection{Energy-Efficient O-RAN Prototype and Simulated RAN Setup}
\label{subsec:experimental-setup}

We developed an experimental environment within the \ac{OSC} platform to implement our energy-efficient solution based on the architecture detailed in Section~\ref{sec:architectural-components}. This environment enables us to simulate realistic \ac{UE} interactions and test our energy-efficient solution under controlled conditions. Previous studies have shown that trends observed at this scale can be reliably extrapolated to larger scenarios via appropriate analytical models and scaling laws~\cite{Baena2021,Zanzi2020}. Consequently, our simulation captures the essential dynamics of energy- and \ac{QoS}-aware load balancing in high-density \ac{RAN} environments, providing valuable proof-of-concept results.

\subsubsection*{Simulation Environment}

To evaluate our solution under realistic conditions with extreme load fluctuations, we configured the simulation to emulate a high-density soccer stadium during a match.
The radio access network comprises 56 \acp{O-RU} placed along the stadium perimeter in a circular pattern with adaptive (nonuniform) spacing.
The number of \acp{O-RU} is comparable in scale to existing works in the literature on network planning for dense stadium environments~\cite{Nakamura2015}.
The deployment density decreases stepwise within each quadrant, starting with a fine angular separation of $2.5^\circ$ in the initial sector, widening to $5^\circ$ in the mid-sector, and extending to $10^\circ$ for the remainder of the quadrant.
\acp{O-RU} are mounted at 25\,m height and set back 5-15\,m from the field edge.
Handover is enabled between any pair of \acp{O-RU}, and each one supports dynamic per-cell power adjustment.

\begin{figure}[t]
  \centering
  \includegraphics[width=.85\linewidth]{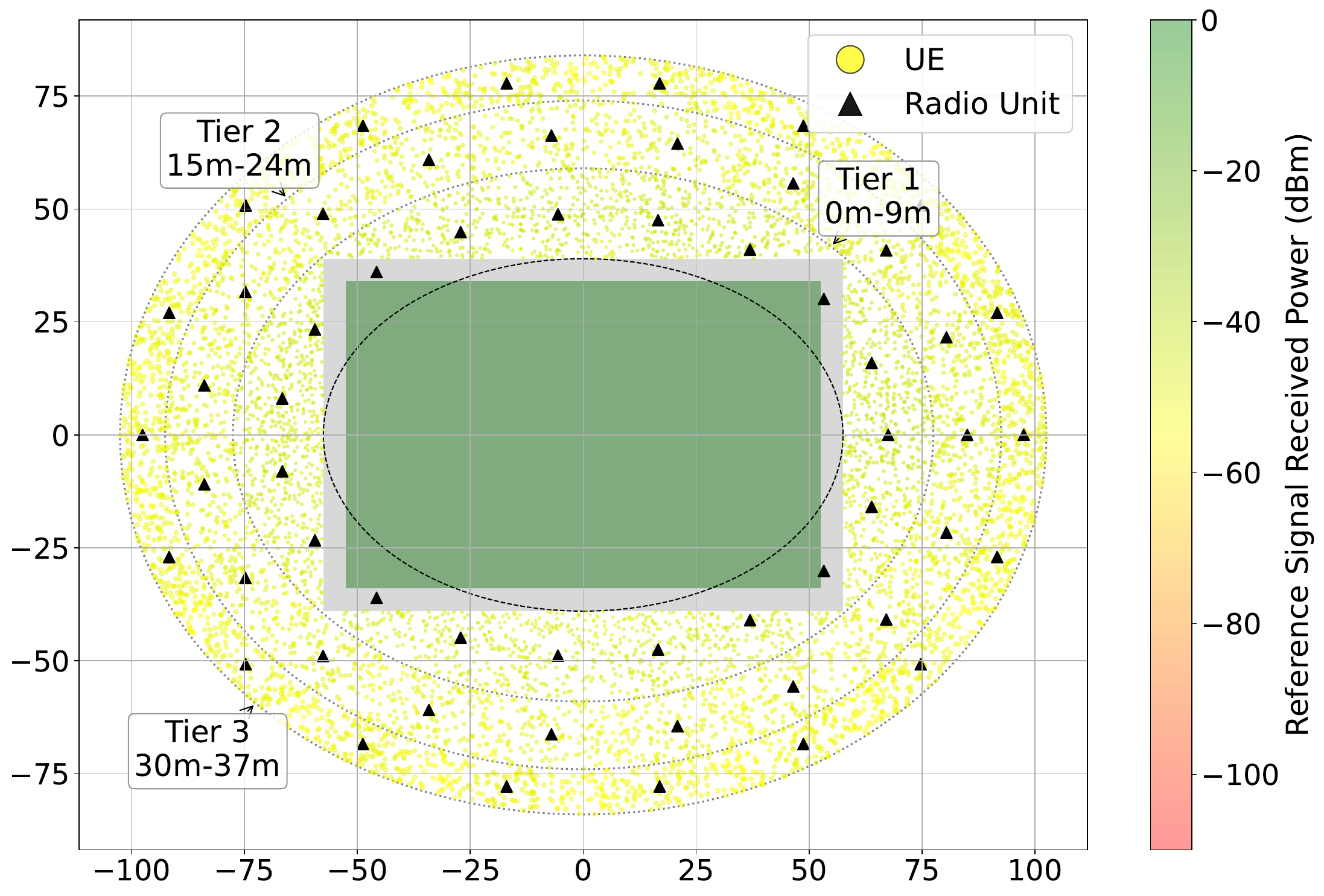}
  \caption{Stadium simulation layout with 56 perimeter O-RUs (triangles) and 8,192 UEs across three tiers; color bar encodes RSRP values. The field dimensions are 105\,m $\times$ 68\,m.}
  \label{fig:stadium-layout}
\end{figure}

The playing field is modeled with the actual dimensions used in our code:  105\,m (length) by 68\,m (width), with a 5\,m technical area around the field.
Fig.~\ref{fig:stadium-layout} depicts the spatial deployment (56 \acp{O-RU}) and \ac{UE} distribution of 8,192 devices, one of the scenarios used in our evaluation, generated using our \ac{RFSS}.

We incorporate the height of \acp{UE} in our simulations to capture the impact of elevation on signal propagation, which is critical in tiered stadium environments where vertical separation affects \ac{LOS} and interference. Grandstands follow a three-tier structure: (i) lower tier at \unit[0]{m} height, \unit[20]{m} depth, $25^\circ$ slope; (ii) middle tier at \unit[15]{m} height, \unit[15]{m} depth, $30^\circ$ slope; and (iii) upper tier at \unit[30]{m} height, \unit[10]{m} depth, $35^\circ$ slope. The \acp{UE} are randomly distributed across all three tiers and are positioned \unit[1.5]{m} above the local tier surface. We summarize all our simulation parameters in Table~\ref{tab:experimental-parameters}.

\subsubsection*{Channel Model}

 We use the \ac{3GPP} Urban Macro (UMa) large-scale model in a simplified close-in form~\cite{3gpp_tr_38_901}:
\[
\mathrm{PL}\,[\mathrm{dB}] \;=\; 28 \;+\; 22\log_{10}(d\,[\mathrm{m}]) \;+\; 20\log_{10}(f_{\mathrm{GHz}}),
\]
with a minimum Tx–Rx separation of 10\,m. Shadow fading is log-normal with $\sigma{=}4.0$\,dB (\ac{LOS}) and $\sigma{=}7.8$\,dB (Non-\ac{LOS}). Based on the received power levels derived from this path loss and fading model, we compute \ac{RSRP}, \ac{RSRQ}, and \ac{SINR} according to the \ac{3GPP} New Radio definitions~\cite{3gpp_ts_38_215}. In line with our implementation, the reference-signal \ac{EPRE} is modeled at 3\,dB below the configured cell transmit power (assumption consistent with typical allocations). Thermal noise density is set to $-174$\,dBm/Hz.

\subsubsection*{Metrics Collection and Traffic Modeling}

During experiments, we collect real-time radio and resource metrics via the \ac{O-RAN} E2 interface using \ac{E2SM-KPM} and \ac{E2SM-RC}~\cite{O-RAN.WG3.E2AP,O-RAN.WG3.E2SM-KPM,O-RAN.WG3.E2SM-RC}. The \ac{E2SM-KPM} provides per-cell resource utilization and per-\ac{UE} \ac{RF} metrics, such as \ac{RSRP}/\ac{RSRQ}/\ac{SINR}~\cite{O-RAN.WG3.E2SM-KPM}, while \ac{E2SM-RC} exposes additional \ac{UE}-level control and mobility information, including \ac{RRC} state, \ac{PCI}, and measurement reports (\ac{RSRP}/\ac{RSRQ}/\ac{SINR}) and registration/handover events~\cite{O-RAN.WG3.E2SM-RC}. Therefore, we model per-\ac{UE} downlink throughput requirements consistent with typical video-streaming usage in crowded sports venues, ranging from approximately 0.5 Mbps for low-resolution highlight clips and social-media interactions to about 25 Mbps for \ac{UHD} live streaming and multi-angle replays, thereby capturing the mix of low- and high-bitrate services that dominates traffic in large stadium environments~\cite{vdocipher2023videobandwidth}.

\begin{table}[t]
\centering
\caption{Simulation Parameters (Stadium Scenario).}
\label{tab:experimental-parameters}
\resizebox{\linewidth}{!}{%
\begin{tabular}{c|c}
\hline
\cellcolor{DarkBlue!80}\textbf{Parameter} &
\cellcolor{DarkBlue!80}\textbf{Value} \\
\hline
\rowcolor{LightBlue!100}
Field Dimensions & \shortstack{105\,m $\times$ 68\,m\\ 5\,m technical area} \\
O-RU Count (Cells) & \shortstack{56 (perimeter) adaptive spacing} \\
\rowcolor{LightBlue!100}
O-RU Height & 25\,m \\
UE Placement & \shortstack{3 tiers; tier-aware depth\\ UE height 1.5\,m above tier} \\
\rowcolor{LightBlue!100}
Carrier Frequencies & \shortstack{3.5--9.0\,GHz (56$\times$100\,MHz)\\ start 3500\,MHz; 100\,MHz steps} \\
Numerology ($n_t$) & 4 \\
\rowcolor{LightBlue!100}
Tx Power ($p_t$) & \shortstack{20\,dBm per O-RU\\ RS EPRE $p_t{-}3$\,dB (assumption)} \\
Antenna Gains ($g_t$, $g_r$) & 8\,dBi (O-RU), 2\,dBi (UE) \\
\rowcolor{LightBlue!100}
Path Loss Model & \shortstack{3GPP UMa (simplified CI):\\ $28{+}22\log_{10}d{+}20\log_{10}f_{\mathrm{GHz}}$; $d_{\min}{=}10$\,m} \\
Shadow Fading $\sigma$ & \shortstack{4.0\,dB (LOS)\\ 7.8\,dB (NLOS)} \\
\rowcolor{LightBlue!100}
Thermal Noise Density & $-174$\,dBm/Hz \\
Max UEs & 9096 provided by Zanzi et~al.~\cite{Zanzi2020}\\
\rowcolor{LightBlue!100}
Handover & Enabled between any O-RUs \\
Power Control & Dynamic per-cell (simulation) \\
\hline
\end{tabular}%
}
\end{table}

Accurate modeling of user arrivals and occupancy is crucial for evaluating energy efficiency in high-density venues, as power consumption is directly linked to the dynamic load of the network.
To capture these match-day dynamics, we incorporate real traces of time series data on user arrivals and occupancy provided by Zanzi et~al.~\cite{Zanzi2020}.
The simulator ingests a time series that controls the gradual addition/removal of \acp{UE}, enabling a faithful representation of stadium occupancy.
It is worth mentioning that we do not consider all \acp{UE} active simultaneously; instead, we assume a realistic traffic usage pattern, discussed further in Section~\ref{subsec:qos-analysis}.

Each \ac{UE} is assigned a fixed position within the stadium stands upon creation, drawn uniformly at random over the angular position along the stadium perimeter and the radial depth within its assigned tier.
This position remains constant throughout the \ac{UE}'s lifetime; no intra-cycle or inter-cycle mobility is modeled.
What varies over time is the \emph{number} of connected \acp{UE}, which follows the time-series input profile and models crowd arrival and departure patterns rather than individual user movement.
Signal metrics (\ac{RSRP}, \ac{RSRQ}, and \ac{SINR}) are therefore computed based on the static Euclidean distance between each \ac{UE} and the \acp{O-RU} using the 3GPP UMa path loss model described above; speed-dependent propagation effects such as Doppler shift and fast fading are not modeled.
This design choice reflects the target scenario: a stadium environment where spectators are predominantly seated and stationary during the event, making individual mobility negligible compared with the crowd-density variation that drives load-balancing decisions.

\subsection{Power Savings - Effective Isotropic Radiated Power}\label{sec:power-eirp}

The \ac{EIRP} metric combines the \ac{RF} power supplied to the antenna and the antenna/beamforming gain into a single comparable quantity. For each cell, we compute its radiated power as the \ac{RF} power supplied to the antenna multiplied by the linear gain (including feeder loss) and sum over the active set to obtain the total $P(t)$. When the actual \ac{RF} power supplied to the antenna or losses are unavailable, we use a normalized version that sets them to 1\,W and 0\,dB. In this context, any constant scale cancels in the savings calculation. We compare our Energy Savings rApp-based optimization to an all-on baseline with all candidate cells active using
\begin{equation}
\eta_{\text{savings}} \;=\; 1 \;-\; \frac{\int_0^T P_{\text{adaptive}}(t)\,dt}{T\,P_{\text{all-on}}}.
\label{eq:savings}
\end{equation}

Because power trajectories share the same multiplicative scale, $\eta_{\text{savings}}$ is the calibration invariant. The resulting value is the fractional reduction in the total transmit-related power relative to the all-on baseline, so larger $\eta_{\text{savings}}$ directly reflects greater progress toward the minimization objective in~\eqref{eq:objective}.

\begin{figure}[ht]
    \centering
    \includegraphics[width=\linewidth]{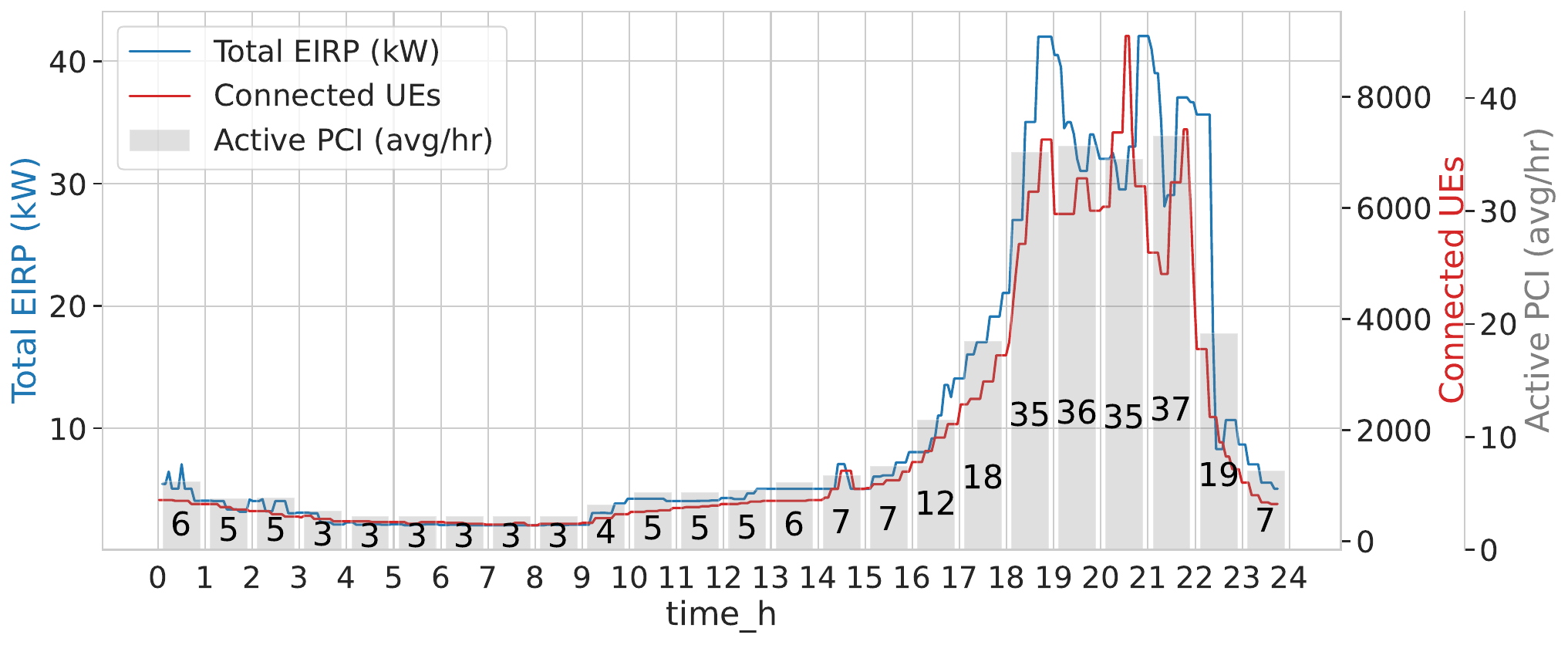}
    \caption{EIRP with the entrance pattern from Zanzi et al.~\cite{Zanzi2020}.}
    \label{fig:eirp-vs-ue}
\end{figure}

To visualize how the Energy Savings rApp decisions manifest across a deployment day, we profile the 24-hour stadium trace under both the adaptive policy and the all-on baseline. Fig.~\ref{fig:eirp-vs-ue} shows a full-day trace, where \ac{EIRP} co-varies with load, while the hourly bars report the mean number of active cells. Peak hours operate near $35$--$37$ active cells and \ac{EIRP} $\approx 30$--$42$\,kW; off-peak hours hold $3$--$6$ cells with \ac{EIRP} $\lesssim5$\,kW. Let $N_{\max}$ denote the total number of candidate cells and let $\mathcal{A}(t)$ be the subset active at time $t$, so $|\mathcal{A}(t)|$ counts active \acp{O-RU} at that instant. We define the instantaneous savings metric as $\eta_{\text{inst}}(t)$. Since $P(t)\propto\sum_i 10^{G_i/10}$ and per-cell gains are comparable within an hour, a conservative bound is
\[\textstyle \eta_{\text{inst}}(t)\;\gtrsim\;1-\frac{|\mathcal{A}(t)|}{N_{\max}}.\]
For example, with $N_{\max}\!=\!56$: at $|\mathcal{A}|\!=\!3$, $\eta_{\text{inst}}\!\approx\!95\%$; at $|\mathcal{A}|\!=\!6$, $\approx\!89\%$; at $|\mathcal{A}|\!=\!12$, $\approx\!79\%$; and at $|\mathcal{A}|\!=\!35$, $\approx\!38\%$. Therefore, our solution achieves substantial savings across most daily intervals by deactivating low-utility cells and judiciously reactivating capacity only as the evening peak approaches the all-on baseline.

\begin{figure}[ht]
    \centering
    \includegraphics[width=\linewidth]{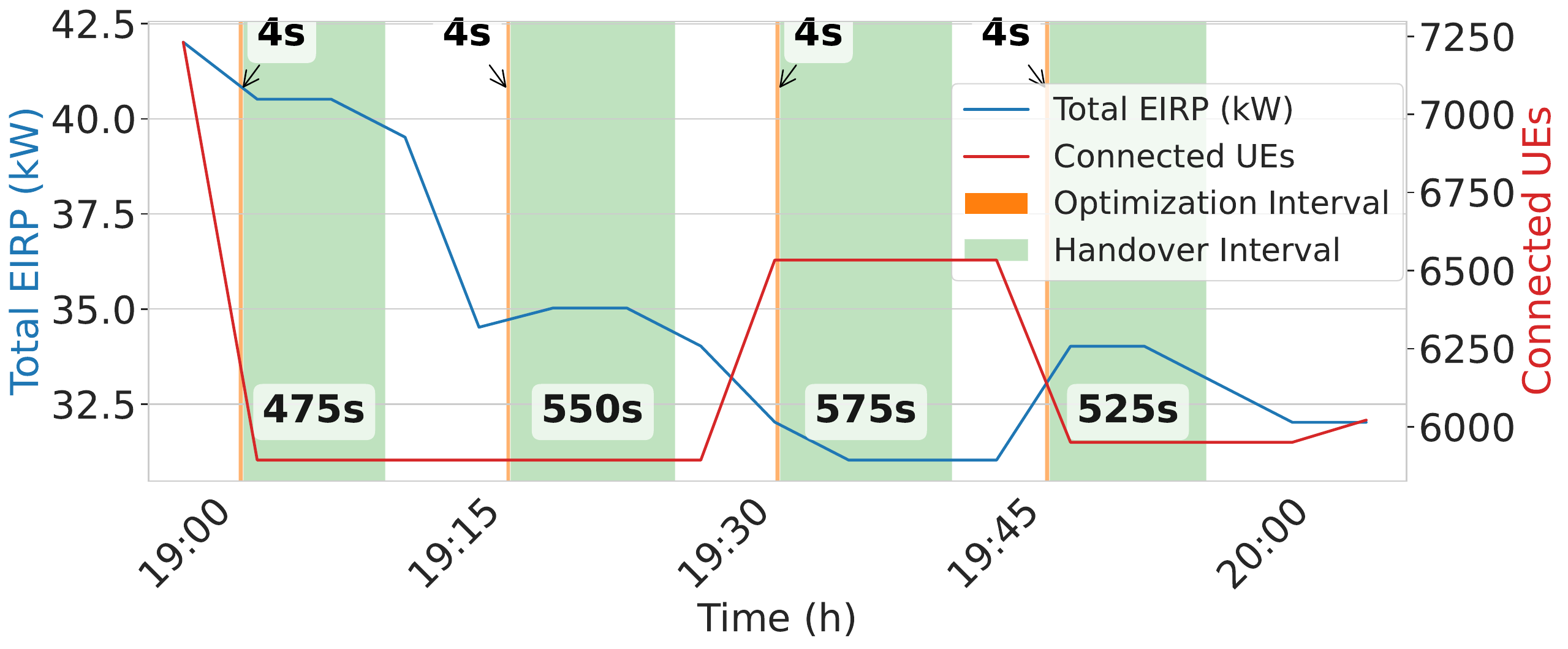}
    \caption{Zoomed 19:00–20:00: short optimization windows (orange) followed by longer handover intervals (green); step-wise EIRP changes with stable plateaus.}
    \label{fig:eirp-vs-ue-zoom}
\end{figure}

To quantify the operational overhead of our Energy Savings rApp and show how little of each cycle is spent actively reconfiguring the \ac{RAN}, we inspect a representative evening peak hour. Fig.~\ref{fig:eirp-vs-ue-zoom} summarizes the timing of Energy Savings rApp actions between 19{:}00 and 20{:}00. The Energy Savings rApp triggers four short optimization windows (orange) of 3.7\,s, 4.2\,s, 4.4\,s, and 4.0\,s (mean $\approx$4.1\,s; range 0.7\,s), each followed by a longer handover interval (green) of 475\,s, 550\,s, 575\,s, and 525\,s (mean $\approx$531\,s; range 100\,s). The average cycle length (optimization $+$ handover) is $\approx$535\,s, with active reconfiguration time occupying only $\sim$0.76\% of the cycle. This $\approx$535\,s cycle falls within the \ac{Non-RT RIC} timescale ($>$1\,s) defined by the O-RAN ALLIANCE, confirming that the Energy Savings rApp operates as a periodic planning loop rather than a near real-time reactive controller (cf.\ Table~\ref{tab:timescale-mapping}). Each decision produces a single step in the \ac{EIRP} trajectory, yielding stable plateaus during the handover intervals and avoiding oscillations or transient overprovisioning, which would otherwise re-trigger \ac{KPI} collection, congest the control plane with churned handovers, and negate energy savings by repeatedly powering cells back on.

The periodic control design inherently provides robustness against cell-state oscillation (ping-pong) and transient dynamics.
Because the Energy Savings rApp aggregates \acp{KPI} over 15\,minute windows, short-lived traffic spikes are smoothed, and the system reacts only to sustained demand trends.
Mid-cycle deviations are corrected at the next cycle boundary ($\sim$535\,s), eliminating the need for intra-cycle reactive adjustments.
This slow control loop acts as implicit hysteresis, ensuring a cell is only (de)activated when warranted by an entire optimization window.
Furthermore, the \ac{EQLite} heuristic is structurally monotone: it greedily adds capacity starting from zero active \acp{O-RU}, meaning a cell is only deactivated during a subsequent full optimization.
Fig.~\ref{fig:eirp-vs-ue-zoom} confirms this: the \ac{EIRP} trajectory exhibits clean, step-wise transitions between cycles without oscillatory reversals, demonstrating stable operation under realistic load evolution.

Power savings arise from two complementary levels visible in the figures: (i) coarse, all-day deactivation of low-utility cells (large off-peak savings, Fig.~\ref{fig:eirp-vs-ue}), and (ii) fine, hour-scale step reductions at nearly constant load (instant $\sim$24\% savings in Fig.~\ref{fig:eirp-vs-ue-zoom}). \eqref{eq:savings} produces a dimensionless quantity, which we call ``scale-free'' because it captures the fractional reduction in \ac{EIRP} regardless of the absolute calibration of transmit power or antenna gain. In this case, the trajectories share the same multiplicative scale that cancels in the ratio. To convert the normalized saving to the actual energy drawn at a site, we add the dominant site-level power components measured on our prototype hardware. These components are ``site-level'' in that they are physical draws present at each cell site independent of traffic normalization: Power Amplifier Direct Current input (648.99\,W), BBU/control electronics (50\,W), and cooling overhead (69.90\,W). Integrating these contributions over the 24-hour trace yields $65{,}440.43$\,Wh ($2.36\times 10^8$\,J) for the all-on baseline and $18{,}267.88$\,Wh ($6.58\times 10^7$\,J) for the adaptive policy, i.e., a $72.08\%$ reduction. The adaptive-day total of 768.89\,W average site power confirms that the scale-free savings materialize as substantial reductions in site energy demand and directly support the minimization goal from~\eqref{eq:objective}.
Our energy-savings evaluation follows the \ac{O-RAN} Network Energy Saving use case~\cite{O-RAN.WG1.NESUC}, which
quantifies energy savings obtained by switching off surplus cells or carrier capacity when traffic demand is low, relative to the corresponding
normal operating conditions.
The $72.08\%$ energy savings refer to this all-on reference.

\subsection{Handover Delay}
\label{subsec:handover-delay}

\begin{figure}[t]
\centering
\includegraphics[width=.85\linewidth]{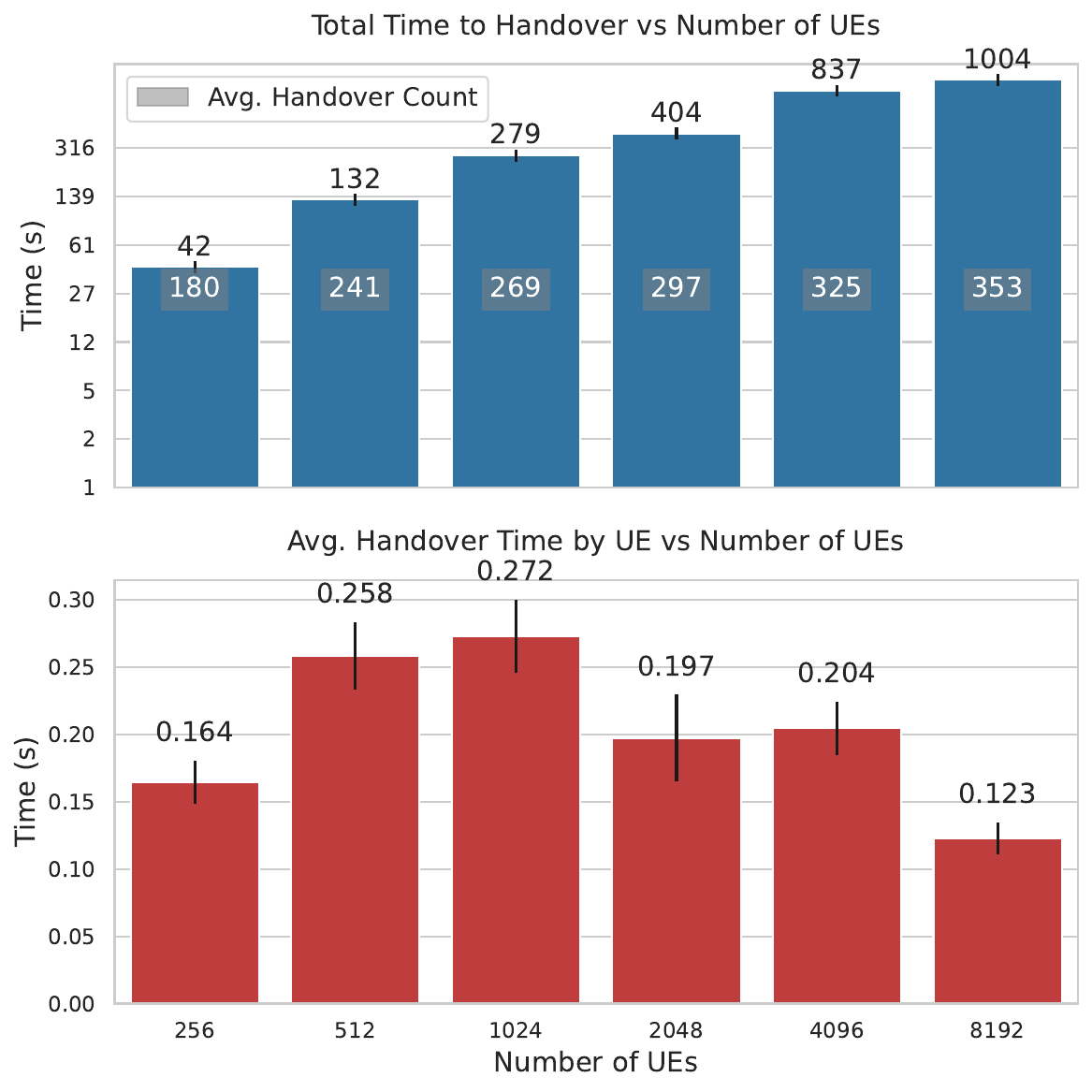}
\caption{Top: total handover time vs.\ number of UEs (log scale; error bars show 10\% st.~dev.; annotated with average handover counts). Bottom: average handover time per UE.}
\label{fig:xApp-handover}
\end{figure}

We assess the handover delay introduced by our \ac{QoS}- and energy-aware load-balancing approach, which assigns \acp{UE} to neighboring cells to ensure service continuity while reducing the number of active E2 Nodes. Starting from an initial state with $56$ active cells, we simulated multiple network loads, ranging from a few hundred to several thousand associated \acp{UE}. 
For each configuration, we measured (i) the total time required to complete all handovers and (ii) the corresponding average delay per \ac{UE}. Fig.~\ref{fig:xApp-handover} presents two complementary perspectives: the top figure shows the aggregate handover time together with the number of executed handovers, while the bottom figure shows our results normalized per \ac{UE}, demonstrating the average handover time in milliseconds. The total handover duration increases with network load, whereas the per-\ac{UE} delay remains sub-second throughout, confirming stable mobility performance under scaling.

Aggregate time increases from $404$\,s at $2048$ \acp{UE} to $837$\,s at $4096$ \acp{UE} ($\approx 0.204\,\mathrm{s}/\ac{UE}$) and to $1004$\,s at $8192$ \acp{UE} ($\approx 0.123\,\mathrm{s}/\ac{UE}$). Two effects explain the behavior under pressure: (a) the energy/\ac{QoS} controller activates additional E2 Nodes under load, reducing per-node contention and suppressing redundant handover triggers via \ac{MLB}-like behavior, and (b) batching/staggering of mobility actions reduces simultaneity, easing transient bursts on the control plane. The small upward drift at $8192$ reflects residual multiplexing and queueing variability at extremely high densities, yet per-\ac{UE} delays remain well below $1$\,s throughout.

It is important to note that the aggregate handover times reported above reflect a sequential, per-\ac{UE} execution workflow in the current prototype: the Handover xApp issues one \ac{E2} \ac{E2SM-RC} control request, waits for confirmation, and only then proceeds to the next \ac{UE}.
This serialization is driven by limitations in the current \ac{E2SM-RC} service model, which does not support batched handover commands and requires individual control transactions for each \ac{UE}.
However, each per-\ac{UE} handover remains an independent \ac{E2} signaling transaction, making the per-\ac{UE} delay (0.123--0.204\,s/\ac{UE}) the architecturally relevant metric, whereas the aggregate time scales linearly ($N \times$\,per-\ac{UE} delay) due to this service model constraint.
With per-cell parallelism, one concurrent worker per \ac{E2} Node, the aggregate time reduces by a factor of $K$ (the number of \ac{E2} Nodes).
For example, at 8{,}192~\acp{UE} with $K{=}56$ cells, the projected aggregate drops from ${\sim}1004$\,s to ${\sim}18$\,s ($1004/56$), bringing the full reconfiguration phase well within a single \ac{Non-RT RIC} cycle.
This optimization requires extending the xApp's concurrency model to issue parallel requests across multiple cells, circumventing the strict sequential limitation without altering the \ac{E2} signaling protocol or the per-\ac{UE} handover semantics.

\subsection{End-to-End Control Delay}
\label{subsec:e2e-delay}

We tested the end-to-end delay of our closed-loop control on a practical \ac{O-RAN} platform under the two monitoring regimes described above (full \ac{VES} pipeline up to 1{,}024~\acp{UE}; direct Prometheus collection beyond) and decomposed each stage's contribution as the \ac{UE} population increased. This metric bounds the frequency with which resources can be reallocated during deployment, helping to identify bottlenecks. We instrumented all stages of one complete control cycle and measured their mean processing time using high-precision timers while varying the number of simulated \acp{UE} from $256$ to $8192$. The stages, whose processing times are detailed in Table~\ref{tab:component-times}, are: Monitoring xApp, the Monitoring System (collection/ingestion pipeline centered on Kafka/\ac{VES}/\ac{VESPA}/Prometheus), the Energy Savings rApp, and the Handover Process. Specifically, Prometheus measures scrape time, the Monitoring System is the Kafka/VES pipeline, the rApp executes the heuristic, and the Handover Process covers control signaling.

To scale large-population experiments, we bypass the monitoring pipeline for runs with more than 1,024 UEs. This bypass is necessary because the ONAP VES Collector's 1\,MB payload limit is exceeded by large cell populations, causing events to be rejected~\cite{onap_ves_spec}, and the VESPA manager's serialization overhead (672\,s at 1,024 UEs) dominates the loop time. Bypassing the VES layer collapses the per-iteration delay to seconds. Utilizing the upcoming ONAP VES High-Volume (VES-HV) interface~\cite{onap_dcae_ves_hv} would resolve this bottleneck.

\begin{table}[t]
\centering
\caption{Component Processing Times per UE Scenario.}
\label{tab:component-times}

\resizebox{\linewidth}{!}{%
\begin{tabular}{c|cccc|c}
  \hline
  \cellcolor{DarkBlue!80}\textbf{\shortstack{UEs}} &
  \cellcolor{DarkBlue!80}\textbf{\shortstack{Prometheus}} &
  \cellcolor{DarkBlue!80}\textbf{\shortstack{Monitoring\\ System}} &
  \cellcolor{DarkBlue!80}\textbf{\shortstack{Energy\\ Savings\\ rApp}} &
  \cellcolor{DarkBlue!80}\textbf{\shortstack{Handover\\ Process}} &
  \cellcolor{DarkBlue!80}\textbf{\shortstack{Total(s)}} \\
  \hline
  \rowcolor{LightBlue!100}
  256  & \shortstack{0.254\\ \footnotesize(0.10\%)} &
          \shortstack{222.16\\ \footnotesize(83.57\%)} &
          \shortstack{1.501\\ \footnotesize(0.56\%)} &
          \shortstack{42\\ \footnotesize(15.79\%)} & 265.915 \\
  512  & \shortstack{0.227\\ \footnotesize(0.04\%)} &
          \shortstack{449.87\\ \footnotesize(77.08\%)} &
          \shortstack{1.515\\ \footnotesize(0.26\%)} &
          \shortstack{132\\ \footnotesize(22.62\%)} & 583.612 \\
  \rowcolor{LightBlue!100}
        1024 & \shortstack{0.420\\ \footnotesize(0.04\%)} &
                                        \cellcolor{red!15}\shortstack{912.66\\ \footnotesize(76.45\%)} &
                                        \shortstack{1.768\\ \footnotesize(0.15\%)} &
                                        \shortstack{279\\ \footnotesize(23.37\%)} & 1193.848 \\
        2048 & \shortstack{0.440\\ \footnotesize(0.11\%)} &
                                        \shortstack{--\\ \footnotesize(-)} &
                                        \shortstack{2.000\\ \footnotesize(0.49\%)} &
                                        \cellcolor{red!15}\shortstack{404\\ \footnotesize(99.40\%)} & 406.440 \\
  \rowcolor{LightBlue!100}
  4096 & \shortstack{0.805\\ \footnotesize(0.10\%)} &
          \shortstack{--\\ \footnotesize(-)} &
          \shortstack{2.950\\ \footnotesize(0.35\%)} &
          \shortstack{837\\ \footnotesize(99.55\%)} & 840.755 \\
  8192 & \shortstack{1.920\\ \footnotesize(0.19\%)} &
          \shortstack{--\\ \footnotesize(-)} &
          \shortstack{5.793\\ \footnotesize(0.57\%)} &
          \shortstack{1004\\ \footnotesize(99.24\%)} & 1011.713 \\
  \hline
\end{tabular}%
}
\par\vspace{2mm}
\footnotesize Times in seconds; percentages are component share of row total. 
\end{table}

Table~\ref{tab:component-times} summarizes the breakdown for $256$–$8192$ \acp{UE}, with the monitoring pipeline enabled (up to $1024$ \acp{UE}). The results reveal two regimes. For up to $1024$ \acp{UE}, the loop is collection/ingestion-bound, with the monitoring pipeline contributing $76$–$84\%$ of the delay. Fine-tuning the scrape cadence, batching strategy, and payload formats, or pruning high-cardinality metrics, would deliver the largest end-to-end gains. Beyond $2048$ \acp{UE}, with direct collection from Prometheus, the loop is execution-bound and dominated by Handover, which motivates parallelizing and batching operations, reducing per-\ac{UE} signaling, and controller-side pipelining to reduce the tail. Across all loads, policy computation in the Energy Savings rApp and Handover xApp is not the limiting factor.

\subsection{Computational Footprint}
\label{subsec:comp-footprint}

\begin{figure}[t]
  \centering
  \includegraphics[width=\linewidth]{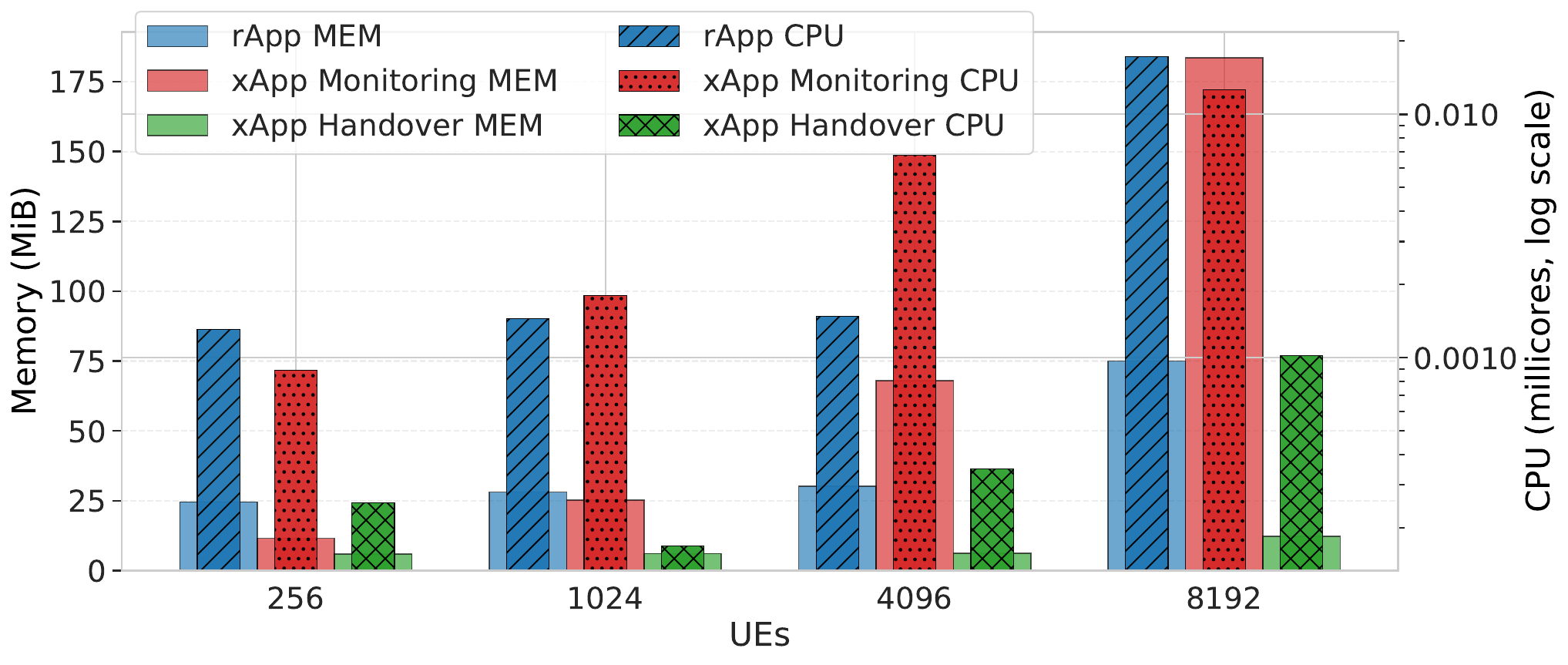}
  \caption{Computational footprint vs. UE density. Solid bars: memory (MiB, left axis). Hatched bars: CPU (millicores, right axis, log scale).}
  \label{fig:comp-footprint}
\end{figure}

We instrumented the CPU (in millicores) and memory (in MiB) of three components in our \ac{O-RAN} closed loop, the Energy Savings rApp, an xApp for continuous monitoring, and an xApp for handovers, while sweeping attached \acp{UE} from $256$ to $8192$. Metrics are collected from container cgroups over fixed windows.
Fig.~\ref{fig:comp-footprint} presents the measured values, where all components scale sublinearly in \ac{UE} count overall and two effects dominate capacity planning: (i) Monitoring memory grows steeply (15.9 times from $256\!\to\!8192$ UEs), becoming the primary memory consumer, and (ii) the Energy Savings rApp shows a high-load knee, with CPU jumping $\sim\!11.7$ times from $4096\!\to\!8192$ UEs. At $8192$ \acp{UE}, the aggregate memory is $\approx270.8$\,MiB ($\approx33.9$\,KiB/UE: $\sim9.4$\,KiB rApp, $23.0$\,KiB monitoring, $1.5$\,KiB handover). These results keep the loop tractable on commodity nodes for up to $8192$ \acp{UE}, while suggesting practical mitigations: batching/downsampling in the monitoring pipeline to reduce memory usage, and Energy Savings rApp sharding or warm-starting to avoid the $8$k–\ac{UE} knee.

\subsection{Throughput Feasibility and QoS Satisfaction}
\label{subsec:qos-analysis}

To assess \ac{QoS} feasibility under the carrier configuration actually used in the simulator, we perform a post-hoc throughput analysis for a reuse-3 deployment in which 56~\acp{O-RU} share three contiguous 100\,MHz carriers at 3.4, 3.5, and 3.6\,GHz. This frequency reuse strategy is standard practice in stadium environments to mitigate severe co-channel interference under pervasive line-of-sight propagation~\cite{SCF270}. The realistic model includes co-channel interference from same-frequency \acp{O-RU} and computes per-\ac{UE} throughput using per-cell fair sharing across simultaneously active users. To capture statistical multiplexing in dense-event traffic, only a fraction of attached users is assumed to compete for PRBs at the same instant, modeled by an activity factor of 0.245 (assuming a 35\% user activity probability and a 70\% downlink \ac{TDD} duty cycle~\cite{Holma2020}). In addition, we estimate the active \ac{O-RU} count per load by aligning the simulated trace with the power log, then interpolating the median active-RU count versus connected \acp{UE}. The nominal 100\,MHz per-\ac{O-RU} bandwidth is then scaled by this load-dependent active-cell ratio. Per-\ac{UE} throughput is therefore computed using an occupancy-calibrated fair-share Shannon model,
$T_{\mathrm{UE}}(N)=n_t \cdot \eta \cdot \frac{\mathrm{BW}_{\mathrm{cell,eff}}(N)}{N_{\mathrm{active}}} \cdot \log_2(1+\mathrm{SINR})$,
where $\mathrm{BW}_{\mathrm{cell,eff}}(N)$ depends on the connected-\ac{UE} load, with practical NR overheads ($\eta=0.70$) and 4x4 MIMO ($n_t=4$).

Table~\ref{tab:qos-satisfaction} reports the percentage of \acp{UE} whose throughput requirements are met in the post-hoc reuse-3 analysis, using effective bandwidth calibrated to the observed cell occupancy.
These additional offline satisfaction percentages complement the optimization as evaluation metrics from our online closed loop shown earlier.
Due to the aggressive energy-saving cell deactivation at low loads, the trace-derived active-RU scaling yields a small effective bandwidth, limiting \ac{QoS} satisfaction to approximately 51--52\% at 256 and 512~\acp{UE}, and 48.1\% at 1{,}024~\acp{UE}.
As load increases, satisfaction does not decline progressively, because the trace-derived bandwidth scaling fluctuates; for instance, satisfaction at 6{,}144~\acp{UE} (38.9\%) is slightly better than at 4{,}096~\acp{UE} (36.2\%).
At higher loads, co-channel interference and per-cell contention dominate: satisfaction falls to 28.5\% at 8{,}192~\acp{UE} and 22.5\% at the observed 9{,}096-\ac{UE} peak.
At 8{,}192~\acp{UE}, achievable throughput averages 7.4\,Mbps with a minimum of 1.4\,Mbps, while the mean \ac{SINR} remains 1.3\,dB.
At 9{,}096~\acp{UE}, mean and minimum throughput fall to 6.0\,Mbps and 1.3\,Mbps, respectively.
These results indicate that the current energy-saving model significantly restricts \ac{QoS} at all load levels, providing roughly 50\% satisfaction at light-to-moderate loads and degrading further under heavy congestion.

\begin{table}[t]
\centering
\caption{Per-UE throughput and QoS satisfaction under reuse-3 deployment with occupancy-calibrated effective bandwidth.}
\label{tab:qos-satisfaction}
\resizebox{\linewidth}{!}{%
\begin{tabular}{r|r|r|r|r|c}
\hline
\rowcolor{DarkBlue!80}
\textbf{\# UEs} & \textbf{Sat.} & \textbf{Mean Tput} & \textbf{Min Tput} & \textbf{Mean SINR} & \textbf{UEs/cell}\\
\rowcolor{DarkBlue!80}
                & \textbf{(\%)} & \textbf{(Mbps)}    & \textbf{(Mbps)}   & \textbf{(dB)}      & \textbf{(attached)}\\
\hline
\rowcolor{LightBlue!100}
   256 &  52.0 & 13.8 & 2.5 & 1.5 &   5.7 \\
   512 &  51.4 & 14.0 & 2.0 & 1.4 &  10.3 \\
\rowcolor{LightBlue!100}
  1024 &  48.1 & 13.8 & 1.3 & 1.2 &  20.2 \\
  2048 &  43.5 & 11.4 & 2.2 & 1.2 &  39.2 \\
\rowcolor{LightBlue!100}
  3072 &  37.0 &  9.6 & 1.6 & 1.2 &  58.2 \\
  4096 &  36.2 &  9.2 & 1.8 & 1.3 &  77.2 \\
\rowcolor{LightBlue!100}
  6144 &  38.9 & 10.0 & 2.1 & 1.3 & 115.7 \\
  8192 &  28.5 &  7.4 & 1.4 & 1.3 & 153.8 \\
\rowcolor{LightBlue!100}
  9096 &  22.5 &  6.0 & 1.3 & 1.3 & 170.6 \\
\hline
\end{tabular}%
}
\end{table}

\begin{figure}[t]
  \centering
  \hfill
  \subfloat[a][Impact of handovers on throughput.]{\includegraphics[width=0.49\columnwidth]{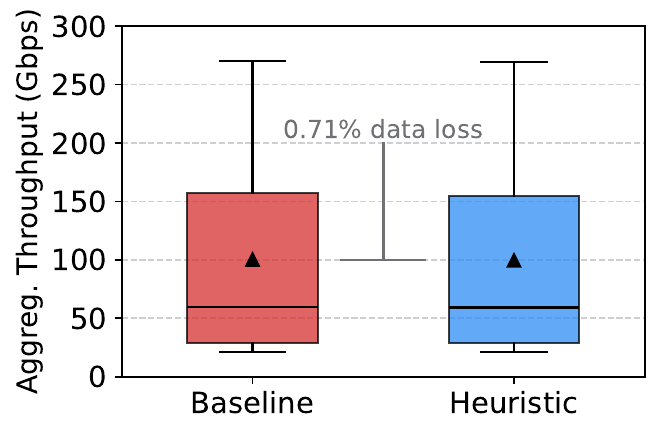}\label{fig:throughput_handover}}
  \hfill
  \subfloat[b][Gain in energy efficiency.]{\includegraphics[width=0.49\columnwidth]{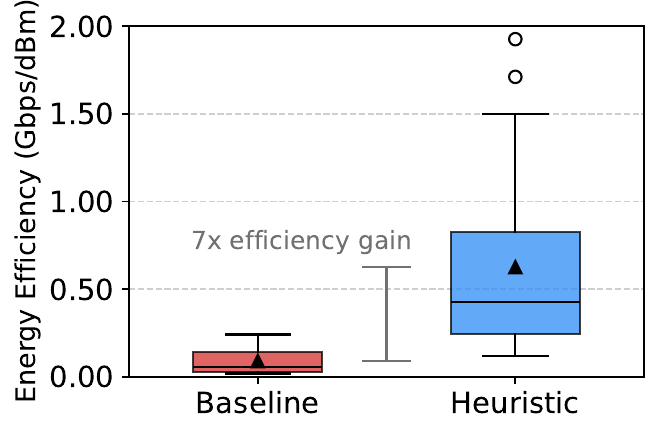}\label{fig:energy_efficiency_handover}}
  \hfill
  \caption{Trade-off between energy efficiency and throughput degradation comparing the baseline and heuristic solutions over the 24-hour evaluation period.}
  \label{fig:handover_impact}
\end{figure}

Beyond static throughput requirements, we must also evaluate service continuity during dynamic cell transitions.
Although 3GPP TS~38.133~\S6.1.3 bounds intra-frequency handover interruption to 30\,ms, we evaluate continuity using an experimentally observed 40.5\,ms interruption window to ensure realistic bounds.
Fig.~\ref{fig:handover_impact} quantifies the impact of handovers on the users' aggregated throughput by comparing our heuristic solution against the baseline approach.
Our heuristic performs more handovers due to dynamic cell activation and deactivation to adapt to temporal traffic variations (see Fig.~\ref{fig:xApp-handover}).
The baseline maintains all cells continuously active to maximize channel quality and reduce data loss.

\begin{figure}[t]
  \centering
  \includegraphics[width=\columnwidth]{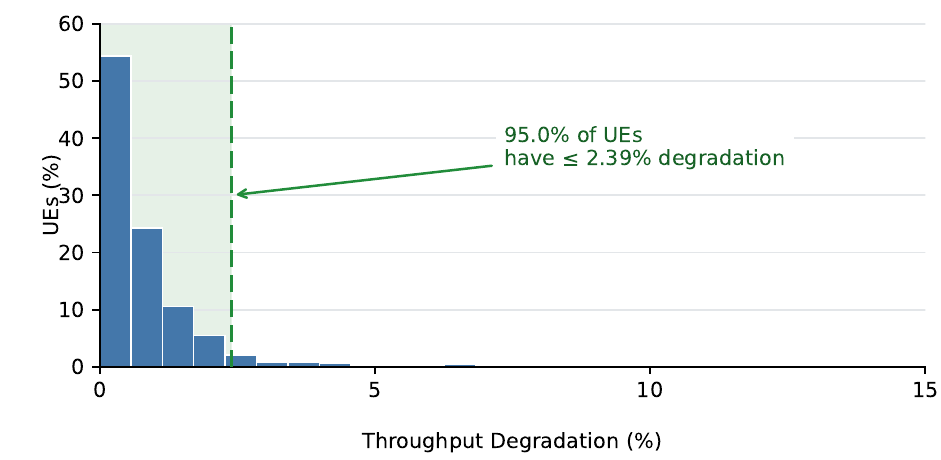}
  \caption{Modeled per-UE throughput degradation relative to the all-on baseline over 24 hours.}
  \label{fig:per-user-throughput-loss}
\end{figure}

The 0.71\% aggregate reduction over 24 hours means that our framework retains approximately 99.29\% of the all-on baseline throughput.
Fig.~\ref{fig:per-user-throughput-loss} further shows that this impact is not uniform across users.
Instead, the degradation follows a narrow distribution: 95\% of UEs experience at most 2.39\% throughput degradation relative to the all-on baseline.
At the same time, our solution achieves approximately 7$\times$ higher energy efficiency by dynamically adapting cell activation and deactivation to demand fluctuations.

%% file: Sections/8-RelatedWorks.tex
\section{Related Work} \label{sec:related-works}

Table~\ref{tab:related-works} presents a year-by-year overview of recent \ac{O-RAN} research that tackles energy efficiency and demand awareness. For each cited work, we list the publication year, reference, and stated goal, and mark seven facets with a three-state symbol: full (\FullCircle), half (\HalfCircle), or empty (\EmptyCircle).  The first five facets locate the contribution within the \ac{O-RAN} stack (xApp, rApp, \ac{Near-RT RIC}, \ac{Non-RT RIC}, and \ac{SMO}). In contrast, the last two, grouped under \emph{Optimization Objectives}, indicate whether the study closes the loop on dynamic-demand control and whether it explicitly reports energy impact.  A {\FullCircle} denotes an implemented and experimentally validated prototype (or, for the Energy objective, measured power-saving evidence), a {\HalfCircle} signals analytical or simulation-only support, and a {\EmptyCircle} shows that the paper does not address the facet.

\begin{table*}[t]
  \centering
  \caption{%
    Qualitative evaluation of related works in O-RAN energy efficiency, dynamic demand, and energy impact.%
  }
  \label{tab:related-works}
  \renewcommand{\arraystretch}{1.15}
  \resizebox{\textwidth}{!}{%
    \begin{tabular}{c ll ccccccc}
      \hline
      \cellcolor{DarkBlue!80}\textbf{Year} &
        \multicolumn{2}{c}{\cellcolor{DarkBlue!80}\textbf{Paper Details}} &
        \multicolumn{5}{c}{\cellcolor{DarkBlue!80}\textbf{O\,-RAN Components}} &
        \multicolumn{2}{c}{\cellcolor{DarkBlue!80}\textbf{Optimization Objectives}} \\ \hline
      \cellcolor{DarkBlue!80} &
        \cellcolor{DarkBlue!80}\textbf{Works} &
        \cellcolor{DarkBlue!80}\textbf{Objective} &
        \cellcolor{DarkBlue!80}xApp &
        \cellcolor{DarkBlue!80}rApp &
        \cellcolor{DarkBlue!80}Near-RT &
        \cellcolor{DarkBlue!80}Non-RT &
        \cellcolor{DarkBlue!80}SMO &
        \cellcolor{DarkBlue!80}Dynamic Demand &
        \cellcolor{DarkBlue!80}Energy \\ \hline
\rowcolor{LightBlue!100}
\textcolor{gray}{2021} & \cite{Ayala-Romero2021} & Mobile video analytics orchestration
  & \EmptyCircle & \EmptyCircle & \EmptyCircle & \EmptyCircle & \EmptyCircle & \FullCircle & \FullCircle \\ \hline
\rowcolor{LightBlue!100}
\textcolor{gray}{2022} & \cite{Abedin2022} & Elastic slicing for \ac{IIoT}
  & \EmptyCircle & \EmptyCircle & \HalfCircle & \HalfCircle & \EmptyCircle & \HalfCircle & \HalfCircle \\
\rowcolor{white}
\textcolor{gray}{2022} & \cite{Hammami2022} & RL-based resource allocation
  & \EmptyCircle & \EmptyCircle & \HalfCircle & \HalfCircle & \EmptyCircle & \HalfCircle & \HalfCircle \\
\rowcolor{LightBlue!100}
\textcolor{gray}{2022} & \cite{Kalntis2022} & vBS performance/energy optimisation
  & \EmptyCircle & \EmptyCircle & \EmptyCircle & \EmptyCircle & \EmptyCircle & \HalfCircle & \HalfCircle \\
\rowcolor{white}
\textcolor{gray}{2022} & \cite{Wang2022} & Computation offloading in IoT
  & \EmptyCircle & \EmptyCircle & \HalfCircle & \HalfCircle & \EmptyCircle & \EmptyCircle & \HalfCircle \\ \hline
\rowcolor{LightBlue!100}
\textcolor{gray}{2023} & \cite{Amiri2023} & Energy-efficient VNF splitting
  & \EmptyCircle & \EmptyCircle & \EmptyCircle & \EmptyCircle & \EmptyCircle & \HalfCircle & \HalfCircle \\
\rowcolor{white}
\textcolor{gray}{2023} & \cite{Mai2023} & Energy efficiency in C-RAN
  & \EmptyCircle & \EmptyCircle & \EmptyCircle & \EmptyCircle & \EmptyCircle & \HalfCircle & \HalfCircle \\
\rowcolor{LightBlue!100}
\textcolor{gray}{2023} & \cite{Salvat2023} & O-RAN testbed design
  & \HalfCircle & \HalfCircle & \FullCircle & \FullCircle & \FullCircle & \FullCircle & \FullCircle \\
\rowcolor{white}
\textcolor{gray}{2023} & \cite{Wang2023} & Energy minimisation in IoT
  & \EmptyCircle & \EmptyCircle & \HalfCircle & \HalfCircle & \HalfCircle & \HalfCircle & \HalfCircle \\
\rowcolor{LightBlue!100}
\textcolor{gray}{2023} & \cite{Ho2023} & Closed-loop energy control
  & \FullCircle & \FullCircle & \HalfCircle & \HalfCircle & \EmptyCircle & \HalfCircle & \FullCircle \\ \hline
\rowcolor{LightBlue!100}
\textcolor{gray}{2024} & \cite{Hoffmann2024} & RF channel reconfiguration
  & \HalfCircle & \HalfCircle & \FullCircle & \FullCircle & \HalfCircle & \FullCircle & \FullCircle \\
\rowcolor{white}
\textcolor{gray}{2024} & \cite{LoSchiavo2024} & Cost/energy in vRANs
  & \HalfCircle & \HalfCircle & \HalfCircle & \HalfCircle & \EmptyCircle & \HalfCircle & \FullCircle \\
\rowcolor{LightBlue!100}
\textcolor{gray}{2024} & \cite{Valcarenghi2024} & Energy-efficient x-haul
  & \HalfCircle & \HalfCircle & \HalfCircle & \HalfCircle & \EmptyCircle & \HalfCircle & \HalfCircle \\
\rowcolor{white}
\textcolor{gray}{2024} & \cite{Hoffmann2024b} & Energy-efficient serving cluster
  & \HalfCircle & \HalfCircle & \HalfCircle & \HalfCircle & \EmptyCircle & \HalfCircle & \HalfCircle \\
\rowcolor{LightBlue!100}
\textcolor{gray}{2024} & \cite{Akman2024} & Multi-vendor cell on/off
  & \FullCircle & \HalfCircle & \FullCircle & \HalfCircle & \HalfCircle & \HalfCircle & \FullCircle \\ 
\rowcolor{white}
  \textcolor{gray}{2024} & \cite{Ros2024} & Priority/demand-based resource management
    & \EmptyCircle & \EmptyCircle & \HalfCircle & \HalfCircle & \EmptyCircle & \HalfCircle & \HalfCircle \\ 
\rowcolor{LightBlue!100}
\textcolor{gray}{2024} & \cite{Bruno2024} & Disaggregated RIC deployment
  & \FullCircle & \EmptyCircle & \FullCircle & \EmptyCircle & \EmptyCircle & \HalfCircle & \EmptyCircle \\ \hline
\rowcolor{LightBlue!100}
\textcolor{gray}{2025} & \cite{Salvat2025} & RU control via ASMs
  & \HalfCircle & \EmptyCircle & \HalfCircle & \EmptyCircle & \EmptyCircle & \HalfCircle & \FullCircle \\
\rowcolor{white}
\textcolor{gray}{2025} & \cite{Lima2025} & Measurement-based power consumption model
  & \EmptyCircle & \EmptyCircle & \FullCircle & \FullCircle & \FullCircle & \FullCircle & \FullCircle \\
\rowcolor{LightBlue!100}
\textcolor{gray}{2025} & \cite{Hidalgo2025} & Energy-efficient computing resource allocation
  & \EmptyCircle & \HalfCircle & \EmptyCircle & \HalfCircle & \HalfCircle & \HalfCircle & \FullCircle \\
\rowcolor{white}
\textcolor{gray}{2025} & \cite{Kim2025OpenRAN} & Network energy saving rApp framework
  & \EmptyCircle & \FullCircle & \EmptyCircle & \FullCircle & \EmptyCircle & \HalfCircle & \HalfCircle \\
\rowcolor{LightBlue!100}
\textcolor{gray}{2025} & \cite{Calagna2025} & Lossless xApp migration \& RIC node scaling
  & \HalfCircle & \FullCircle & \FullCircle & \FullCircle & \EmptyCircle
  & \FullCircle & \FullCircle \\ \hline

\rowcolor{DarkBlue!80}
 & \textbf{This Work} & Energy-aware dynamic resource management
  & \FullCircle & \FullCircle & \FullCircle & \FullCircle & \FullCircle & \FullCircle & \FullCircle \\ \hline
    \end{tabular}}
  \footnotesize
  \FullCircle\,measured / experimentally validated\enspace
  \HalfCircle\,analytical or simulation only\enspace
  \EmptyCircle\,not addressed
\end{table*}

Integrating adaptive solutions into 5G Advanced \ac{O-RAN} platforms is crucial for telecommunication networks.
State-of-the-art literature has extensively reviewed the energy efficiency aspects of \ac{O-RAN}~\cite{Abubakar2023, Rimedo2023, O-RAN.WG1.NESUC}.
Recent experimental evaluations include the first measurement-based analysis of \ac{RIC} power consumption overheads~\cite{Lima2025}.

Various resource management and optimization frameworks have been proposed for disaggregated RAN architectures, addressing video analytics~\cite{Ayala-Romero2021}, elastic slicing~\cite{Abedin2022}, reinforcement learning~\cite{Hammami2022}, performance optimization~\cite{Kalntis2022}, computation offloading~\cite{Wang2022}, VNF splitting~\cite{Amiri2023}, Cloud-RAN energy efficiency~\cite{Mai2023}, testbeds~\cite{Salvat2023}, and closed-loop control~\cite{Ho2023}.
On the radio side, studies focus on RF channel reconfiguration~\cite{Hoffmann2024}, cost-aware design~\cite{LoSchiavo2024}, x-haul routing~\cite{Valcarenghi2024}, cell-free MIMO~\cite{Hoffmann2024b}, multi-vendor coordination~\cite{Akman2024}, sleep-state control~\cite{Salvat2025}, computing resource allocation~\cite{Hidalgo2025}, and MEC resource management~\cite{Ros2024}.

Additionally, several software-development frameworks have emerged to support green O-RAN control loops.
For instance, Kim et al.~\cite{Kim2025OpenRAN} present an open-source development and visualization framework for energy-saving rApps on the Non-RT RIC.
Calagna et al.~\cite{Calagna2025} propose CORMO-RAN to migrate stateful xApps and power down idle controllers, achieving up to 64\% energy savings.
Further, Zanatta Bruno et al.~\cite{Bruno2024} evaluate the deployment of disaggregated Open RAN controllers on distributed clouds.

Our work provides practical validation through an \ac{O-RAN} testbed integrating the \ac{Non-RT RIC}, the \ac{Near-RT RIC}, and the \ac{SMO}.
While most related work relies primarily on simulation and recent experimental studies focus on specific components, our approach uniquely implements a complete control loop spanning the entire \ac{O-RAN} system.
The testbed validates functional correctness and control-plane integration at all tested scales (up to 8{,}192~\acp{UE}), while demonstrating full observability-pipeline scalability (E2/VES/VESPA/Kafka) up to 1{,}024~\acp{UE}; beyond that threshold, a simplified Prometheus-based monitoring path is used (see Section~\ref{subsec:e2e-delay}).
By combining the global perspective of the \ac{Non-RT RIC} with the local control capabilities of the \ac{Near-RT RIC}, and further enhanced by the \ac{SMO}, our solution bridges the gap between simulation studies and effective experimental validations.

%% file: Sections/9-Conclusion.tex
\section{Conclusion} \label{sec:conclusion}

This article addresses the critical need for energy efficiency in 5G Advanced networks by introducing an adaptive network management solution designed for \ac{O-RAN} architectures. The solution provides a hierarchical control framework in which the \ac{Non-RT RIC} periodically computes energy-aware policies from aggregated \acp{KPI}, while the \ac{Near-RT RIC} enforces those policies through per-\ac{UE} handover actions on a sub-second timescale. This design optimizes the allocation of \acp{UE} and the transmission power of radio base stations, ensuring scalability for high-density environments such as stadiums. Key contributions include the development of an adaptive management solution aligned with \ac{O-RAN} standards and the introduction of novel energy-efficiency techniques.
Empirical validation confirmed that handover delays remained subsecond (0.123--0.204\,s/UE) across 256--8192~\acp{UE} without coverage gaps. The adaptive control policy achieved 72.08\% energy savings over a 24-hour trace (from $\approx$65.4\,kWh to $\approx$18.3\,kWh), maintaining 3--6 active cells during off-peak hours, compared with $\sim$35--37 during peak load. However, these savings come at the cost of \ac{QoS}; aggressive cell deactivation yields roughly 50\% throughput satisfaction at low-to-moderate loads, and degrades further under extreme congestion. Future work will extend this solution to realistic mobility scenarios, including pedestrian-speed mobility models (e.g., random waypoint or group mobility) to simulate spectator movement during intermissions and ingress/egress phases, speed-dependent signal degradation effects such as Doppler shift and fast fading, and evaluation of the proposed load-balancing solution under varying mobility speeds to assess its robustness to mobility-induced handovers and signal fluctuations, as well as parallelized handover execution for reduced aggregate reconfiguration time, migration to standardized A1-TD policy types as they mature to cover per-UE reassociation semantics, full-pipeline \ac{VES}\slash{}\ac{VESPA} scalability beyond 1{,}024~\acp{UE} enabled by the upcoming ONAP \ac{VES} High-Volume interface, additional \ac{QoS} indicators such as latency, jitter, and outage probability, as well as \ac{AI}/\ac{ML}-driven adaptability to dynamic traffic conditions.

%% file: Sections/10-Acknowledgment.tex
\section*{Acknowledgment}
This work was partially supported by CNPq Grants Nos.\ 405111/2021-5 and 130555/2019-3, and by CAPES, Finance Code 001, Brazil; additional support was provided by RNP and MCTIC under Grant No.\ 01245.010604/2020-14 as part of the 6G Brasil and OpenRAN@Brasil projects, and by MCTIC/CGI.br/FAPESP through Project SAMURAI (Grant No.\ 2020/05127-2) and Project PORVIR-5G (Grants No.\ 2020/05182-3 and 2025/01970-0); it has also been fully/partially funded by the project XGM-AFCCT-2024-5-1-1 supported by xGMobile - EMBRAPII - Inatel Competence Center on 5G and B5G Networks, with financial resources from the PPI IoT/Manufatura 4.0/the MCTI grant number 052/2023, signed with EMBRAPII; and finally, this work also received support from the Commonwealth Cyber Initiative (www.cyberinitiative.org).


%% file: Sections/11-Biography.tex
\begin{IEEEbiographynophoto}{Gustavo Z. Bruno}
is a postdoctoral researcher at INATEL and an IT infrastructure analyst at MTI. His research interests include infrastructure automation, O-RAN, and cloud computing.
\end{IEEEbiographynophoto}

\begin{IEEEbiographynophoto}{Gabriel M. Almeida}
is a Ph.D. candidate in Computer Science at UFG. His research interests span wireless networks, virtualization, and resource allocation.
\end{IEEEbiographynophoto}

\begin{IEEEbiographynophoto}{Aloízio Da Silva}
is the xG Testbed Director at CCI, Virginia Tech, and faculty in Electrical and Computer Engineering. His research focuses on network softwarization and testbed management.
\end{IEEEbiographynophoto}

\begin{IEEEbiographynophoto}{Luiz A. DaSilva}
(Fellow, IEEE) is the Executive Director of the Commonwealth Cyber Initiative (CCI) and Bradley Professor at Virginia Tech. His research focuses on adaptive resource management in wireless networks.
\end{IEEEbiographynophoto}

\begin{IEEEbiographynophoto}{Joao F. Santos}
(Member, IEEE) is a research assistant professor at CCI, Virginia Tech. His research interests include open radio access networks and software-defined wireless networks.
\end{IEEEbiographynophoto}

\begin{IEEEbiographynophoto}{Alexandre Huff}
is an associate professor at UTFPR, Brazil. His research interests include distributed systems, fault tolerance, and open radio access networks.
\end{IEEEbiographynophoto}

\begin{IEEEbiographynophoto}{Kleber V. Cardoso}
is an associate professor at UFG, Brazil. His research focuses on wireless networks, virtualization, and performance evaluation.
\end{IEEEbiographynophoto}

\begin{IEEEbiographynophoto}{Cristiano B. Both}
is a professor at UNISINOS, Brazil. His research interests include wireless networks, softwarization, and virtualization technologies.
\end{IEEEbiographynophoto}